\documentclass[manuscript]{acmart}
\usepackage{multirow}
\usepackage{array}

\AtBeginDocument{%
  }

\begin{document}

\title[Surveillance and Disability in Online Proctored Exams]{Surveillance and Disability in Online Proctored Exams: Student Perspectives and Design Implications}

\author{Monika Blue Kwapisz}
\affiliation{%
  \institution{Industrial and Systems Engineering, University of Washington}
  \city{Seattle}
  \state{WA}
  \country{USA}
}

\author{Yoav Ackerman}
\affiliation{%
  \institution{Industrial and Systems Engineering, University of Washington}
  \city{Seattle}
  \state{WA}
  \country{USA}
}

\author{Jennifer Nguyen}
\affiliation{%
  \institution{Industrial and Systems Engineering, University of Washington}
  \city{Seattle}
  \state{WA}
  \country{USA}
}

\author{Dr. Prashanth Rajivan}
\affiliation{%
  \institution{Industrial and Systems Engineering, University of Washington}
  \city{Seattle}
  \state{WA}
  \country{USA}
}



\begin{abstract}
  Online proctoring systems (OPS) are technologies and services that are used to monitor students during an online exam to deter cheating. However, OPS often violates student privacy by implementing overly intrusive surveillance to which students cannot consent meaningfully. The technologies used in OPS have been shown to unfairly flag students with disabilities. Our reflexive thematic analysis of interviews with students who have first-hand experience with online invigilated exams and who have disability accommodations points to their anxiety about the interaction between surveillance and their disabilities, leading to fears about misrepresentation and increased cognitive load on the exam. Students describe the compromises they need to make with their privacy and accommodations to take remote tests and share their privacy values. We present the implications for the design of OPS to mitigate the issues faced by disabled students.
\end{abstract}

\maketitle

\section{Introduction}

Proctoring (as it is referred to in the United States) or invigilation (as it is referred to in the UK and other English-speaking countries) is the monitoring of students during an exam to deter them from cheating~\cite{cambridge_university_press_proctor_nodate}. Online proctoring systems (OPS) are a variety of technologies and services used to implement proctoring remotely. Examples of situations where OPS are usually used include entrance exams to academic programs, proficiency exams for international students, and Massive Open Online Courses (MOOCS)~\cite{ras_online_2018}. In the past, MOOCS used testing centers with in-person proctors and live proctors monitoring students through an OPS, but in the last ten years, innovation in OPS has expanded to a variety of automated methods of proctoring~\cite{li_massive_2015,li_visual_2021}. Emergency remote teaching (ERT) necessitated by the COVID-19 pandemic caused an explosion in the use of OPS beyond MOOCS. For example, OPS was implemented in higher education institutions (HEIs) without paying adequate attention to the privacy implications of the newly adopted technology~\cite{khalil_nexus_2022,selwyn_necessary_2023}. Education technology companies benefit from the growth of this technology adoption~\cite{harwell_cheating-detection_2020} while failing to sufficiently acknowledge the effects of surveillance on student privacy and well-being~\cite{selwyn_necessary_2023,coghlan_good_2021,fawns_matter_2022}. Privacy violations aren't the only issue with OPS: popular media~\cite{caplan-bricker_is_2021, feathers_proctorio_2021,hill_accused_2022,patil_how_2020} and academic literature~\cite{burgess_watching_2022,johanson_lockdown_2025} have discovered that OPS discriminate against users of color and disabled users. Disabled OPS users face false suspicions and accusations of cheating based on behaviors such as stimming~\cite{cambridge_advanced_learners_dictionary__thesaurus_stimming_2013}, fidgeting, and using the restroom during an exam. These behaviors are exacerbated by anxiety~\cite{lin_looking_2023,kapp_people_2019,hernandez_influence_2021}. Our paper investigates the aspects of OPS surveillance that students with disabilities say make their anxiety worse, and how that anxiety creates a cycle of misrepresentation and added cognitive load during testing. An unnecessary increase in cognitive load is troublesome for all students, but many students with disability accommodations have psychological and learning disabilities that make the increased cognitive load unacceptably burdensome. We investigate how privacy-violating surveillance in OPS contributes to discrimination and ableism through the following questions:

\subsection{Research questions}

\begin{enumerate}
    \item How do perceptions of OPS surveillance impact testing anxiety among students with disabilities?
    \item How does surveillance anxiety impact disabled students' behavior and cognitive load during exams?
    \item How do individual differences in lived experiences of disability affect students' behaviors and perceptions toward being misrepresented by OPS surveillance?
    \item What values do students with disabilities have about their privacy while interacting with OPS?
\end{enumerate}

\subsection{Contributions}

We present quotes from sixteen semi-structured interviews that highlight the various lived experiences of taking OPS exams with disability accommodations. Privacy and surveillance are usually briefly discussed by researchers studying discrimination based on disability in OPS and education technology~\cite{gin_covid-19_2021,pierres_artificial_2023,carter_attitudes_2024,johanson_lockdown_2025}. Our paper offers the first collection of multiple and diverse disabled students' perceptions and attitudes about surveillance. We propose the following design interventions to mitigate commonly faced issues.

\begin{enumerate}
    \item Use contextual integrity to address asymmetric information sharing by designing for transparency, appropriateness, and consent.
    \item Address anxiety with ability-based design, for example, build calming routines into OPS that would benefit anxious students regardless of whether they are disabled or able-bodied. 
    \item Starting the OPS design process from universal design in higher education (UDHD) principles and privacy by design (PbD) to build accessibility and privacy into the system from the beginning, rather than retroactively remedying surveillance harms.
\end{enumerate}

\section{Background}

\subsection{OPS privacy issues} 

OPS can be divided into three types: live proctoring, where a human proctor watches a student in real time, recorded proctoring, where cameras and logs are reviewed later, and automated proctoring, where the proctoring system identifies suspicious events for review by a human proctor~\cite{hussein_evaluation_2020,arno_state---art_2021}. Common features are eye movement tracking, facial recognition, object and face detection, and log analysis~\cite{hussein_evaluation_2020}. Usually, the student's computer webcam and microphone provide the input for various algorithms designed to detect anomalies, which researchers claim indicate cheating~\cite{atoum_automated_2017,dilini_cheating_2021}. Sometimes students take online exams at testing centers, so we also consider in-person, human proctors and cameras installed in the testing rooms as part of the broader experience of using an OPS. When the OPS determines that the student is out of compliance or detects something suspicious, it ``flags'' the student. 

Researchers have studied how OPS surveillance harms students' privacy. Burgess et al. identify two major student privacy concerns: informed consent and post-exam monitoring. The OPS Burgess et al. studied had terms and conditions that either did not inform students about the specific data being collected or offered conflicting information. Disturbingly, one OPS in Burgess et al.'s study constantly ran on the computer, logging applications and installing updates, after a student was no longer taking an exam~\cite{burgess_watching_2022}. Terpstra et al. find that a variety of the information collected and shared is unacceptable to students, especially video recordings of their personal spaces and personal information~\cite{terpstra_online_2023}. Although OPS are required to comply with privacy regulations~\cite{slusky_cybersecurity_2020}, legal scholars point out that OPS may violate General Data Protection Regulation (GDPR) largely due to students being deprived of the opportunity to meaningful consent to the data being shared~\cite{barrio_legal_2022}. Woldeab has documented that students have a ``fear of proctor intrusiveness''~\cite{woldeab_under_2017} and that further research is needed into the effect of the presence of a live proctor on high anxiety students~\cite{woldeab_21st_2019}. Other research has also shown that students find proctoring invasive~\cite{ul_haque_nuanced_2023,dimeo_online_2017}. Ul Haq et al. documented a feeling of injustice where a participant called the OPS ``cruel and unusal''~\cite{ul_haque_nuanced_2023}.

\subsection{OPS discrimination}

Journalists have brought attention to the racism encoded in the facial detection technologies used by OPS that may cause more flags of Black people~\cite{feathers_proctorio_2021}, especially Black women and girls~\cite{hill_accused_2022}. Black and Afro-Latina students have reported shining bright lights on their faces to have them detected by the OPS, which causes discomfort and disruption to the testing experience~\cite{caplan-bricker_is_2021,patil_how_2020}. Academics agree that OPS are more likely to flag students with darker skin tones as suspicious~\cite{yoder-himes_racial_2022,pilgrim_online_2024}.

A common problematic OPS feature is flagging students who look away from the screen or have unusual movements, such as neurodivergent students who stim~\cite{pilgrim_online_2024}. Researchers have found that this feature is ``discriminatory against a parent who attends to a young child, a person who cares for an elder, and anyone who cannot look straight at a screen for long periods of time due to a medical condition''~\cite{barrio_legal_2022}. For example, the New York Times interviewed a student who became a caregiver to her younger siblings when the COVID-19 pandemic necessitated at-home testing; the OPS flagged loud sounds made by the children, causing the student to feel a lack of belonging~\cite{patil_how_2020}.

Disabled students using their accommodations are also flagged by OPS, like one student who was ejected from a test for using a zooming function, according to a story in the New Yorker~\cite{caplan-bricker_is_2021}. Another student recounted how the OPS was not made aware of their accommodations, which caused a delay in the test~\cite{patil_how_2020}. Students with accommodations may also have more trouble installing and using OPS software on personal devices due to a lack of universal accessibility in OPS design~\cite{carter_attitudes_2024}. Meanwhile, other disabled students cannot care for their needs due to the restrictions on movement, making it so they couldn't take breaks to go to the bathroom~\cite{gin_covid-19_2021} or leave the room for medical reasons~\cite{johanson_lockdown_2025,gin_covid-19_2021}.

\subsection{Panopticon surveillance theory} \label{panopticon}

Surveillance is commonly understood through the metaphor of the panopticon, written about by Michel Foucault~\cite{foucault_discipline_1977}. The panopticon is a prison made up of individual cells arranged cylindrically around a watchtower so that the watcher can potentially see all the cells at once. The prisoners can not see each other, and they cannot see the watcher due to a bright light on the watchtower. The prisoner feels constantly surveilled due to the uncertainty of when the watcher is watching (if at all)~\cite{foucault_discipline_1977}. The uncertainty of whether a proctor is watching and the one-sidedness of the proctoring are two ways in which OPS mirrors the panopticon. Other researchers have also compared OPS to the panopticon~\cite{peixoto_when_2022}. A major issue with panoptic surveillance is that it has a chilling effect that has been shown to make people more cautious in their online activities and less willing to dissent~\cite{stoycheff_privacy_2019}. This is seen as a positive by some researchers in the OPS context because it may reduce cheating on online exams~\cite{hylton_utilizing_2016,slusky_cybersecurity_2020}. However, the data is not conclusive about whether OPS offer more than the placebo effect for cheating prevention~\cite{conijn_fear_2022,bergmans_efficacy_2021}, which calls into question the ethics of surveillance without a clear justification, especially considering that surveillance harms intellectual activity~\cite{grayson_education_1978,richards_intellectual_2008}.

\subsection{Disability theory} \label{disability}

The Americans with Disabilities Act (ADA) requires that testing entities provide accommodations in a timely manner to students who provide reasonable documentation that they have an impairment that substantially limits a major life activity so that tests ``accurately reflect the individual’s aptitude or achievement level''~\cite{us_department_of_justice_civil_rights_division_ada_2020}. The participants in this study are all students who self-report having these types of official accommodations during testing; however, their identities as disabled students differ. Many students who use disability accommodations identify as having a ``hidden'' disability (learning, cognitive, or psychiatric condition), which may lead them not to identify as ``disabled'' or not disclose their disability status~\cite{edwards_academic_2022,melian_getting_2022}. Additionally, neurodiversity proponents reframe certain cognitive differences as forms of diversity rather than impairments~\cite{botha_neurodiversity_2024, singer_neurodiversity_2017}. In this article, we will refer to the broad population of students with various disability testing accommodations as ``students with disabilities'' or ``disabled students,'' while keeping in mind that their actual identities may differ.

We write from the perspective of the social model of disability, which defines disability according to the societal and environmental barriers for people with disabilities rather than their individual impairments~\cite{oliver_politics_1990}. However, other scholars have argued for balancing the social model of disability with the difficulties faced inherently due to impairments~\cite{shakespeare_social_2006}. The social model of disability is useful for us as human-computer interaction researchers because we seek to redesign systems where there are barriers to access~\cite{armagno_role_2012}.

\section{Methods}

\subsection{Research ethics}

We upheld strong research ethics by protecting participants from harm, protecting their data, and ensuring high standards among our research team. This study was classified as exempt by the authors' university's Institutional Review Board (IRB) because it was determined that the study poses no more than minimal risk to participants. We exceeded the requirements of an IRB-exempt study by providing participants with a consent form that detailed potential discomfort during the study and who to contact in the case of harm due to the research. We checked for understanding of the consent form and the aims of the study at the beginning of each interview. We ensured that the participants remained anonymous through a pseudonym. Pseudonyms are stored separately from the real names of the participants. The interview recordings will be deleted at the end of this study. All members of the research team have undergone training in human subjects research ethics.

\subsection{Interviews} \label{interviews}

We used a semi-structured, intensive interviewing process as described by Charmaz~\cite{charmaz_constructing_2006} because we wanted to explore the emotional depth of students' experiences using OPS. The interview guide was created based on reviewing literature about mental models of privacy~\cite{oates_turtles_2018,kang_my_2015, pflugfelder_reddits_2017},  misrepresentation on online educational tools~\cite{kwapisz_privacy_2024, slade_learning_2013,depew_mediating_2009,webber_limitations_2019}, methods for cheating prevention using OPS~\cite{atoum_automated_2017, dendir_cheating_2020}, and ableism in OPS~\cite{caplan-bricker_is_2021,patil_how_2020,carter_attitudes_2024,gin_covid-19_2021,johanson_lockdown_2025}. The topics and example questions can be found in Appendix~\ref{guide}. The interview guide was tested using five pilot interviews to refine the interview protocol. The inclusion criteria were: 
\begin{enumerate}
    \item Being over eighteen years of age.
    \item Being a part-time or full-time student at a college or university in the United States.
    \item Having official accommodations via the campus disability resource office that are used when taking tests.
    \item Using an OPS in the last two years from the interview date, which was defined as a test with a lockdown browser, AI proctoring, a webcam recording during the exam, the computer screen being recorded during the exam, or an online proctor monitoring the exam.
\end{enumerate} 
Potential participants responded to a screening survey on Qualtrics to determine their eligibility (based on the inclusion criteria) to participate in the study. Participants who were determined eligible based on the screening survey were interviewed by the first author using Zoom. The participants were paid between \$15 and \$20 for a one-hour interview, provided either as gift cards (for those recruited through offline channels) or through Prolific payments (for those recruited via Prolific). Participants responded to a demographic questionnaire after the interview. The results can be found in Appendix~\ref{demographics}.

We used a combination of approaches to recruit study participants. We started by recruiting at the authors' university by emailing a flyer to students through their academic advisors, posting flyers at the disability resource testing center, and social media posts. Five participants were recruited at the authors' HEI using these methods. As we analyzed these interviews, we realized the study would benefit from a more diverse sample of experiences, so we used Prolific to recruit thirteen more participants. Two participants who were included based on the screening survey were later found ineligible during the interview, consistent with previous findings on the issue of inattentive survey participants on Prolific~\cite{douglas_data_2023}. We analyzed sixteen interview transcripts.

\subsection{Analysis}

We used a reflexive thematic analysis method~\cite{braun_thematic_2022} to synthesize the transcripts of the interviews into three major themes. Interviewing and analysis happened concurrently over the course of seven months. First, the research team used clean verbatim guidelines to edit the Zoom-generated transcripts of our interviews to be easily readable. Second, we structural coded the transcripts to scope this article on the inclusion criteria of participants talking about privacy and surveillance and experiences with OPS. Exclusion criteria were rapport-building, experiences with testing not using an OPS, non-exam conversations about surveillance, and discussions about cheating prevention. Third, the team individually coded the selected portions of the interviews in sets of about two participants based on their interpretations of the participants' statements using NVivo. Although Braun and Clarke recommend against consensus coding~\cite{braun_thematic_2022}, a consensus-based peer data analysis approach was taken to generate ideas for creating new codes and refining their definitions because it was useful in prompting meaningful discussion between the group. Once the codebook had clear definitions for each code and made up a substantive basis for theoretical analysis, further interviewing and coding were stopped. This is consistent with Braun and Clarke's views on stopping data collection and analysis based on reflexive understandings of the work rather than positivist framings~\cite{braun_saturate_2021}. The team wrote reflexive memos throughout the analysis process about how they were approaching coding, the connections between codes, what themes they saw emerging, and their personal connections to the work. These reflexive memos became central to the theme generation phase, where we compiled the codebook, the contents of the codes, and preliminary ideas from the memos to iteratively explore how the codes connected and differed, in addition to brainstorming and reflecting with the team. 

\subsection{Limitations}

We consider reflexive thematic analysis to be the most appropriate analysis method for our work because it is based on the social construction of knowledge~\cite{braun_thematic_2022}. However, a positivist researcher might find it limiting that we cannot describe our research as offering objective truth on the experience of disability and surveillance on OPS. We also can't generalize our work because the lived experiences of students with disabilities differ, and we can't present every experience. We are also limited in the diversity of the disabilities, identities, and other demographics of our participants because we did not intentionally select for any demographic beyond those listed in our inclusion criteria in Section~\ref{interviews}. 

\section{Results} \label{results}

Three themes were generated based on analysis of the codebook and definitions. The themes reflect underlying commonalities in disabled students' experiences while highlighting their different lived experiences through quotes found under each theme subsection.

\subsection{Negative emotions, like anxiety, lead to fear of misrepresentation and increased cognitive load while taking a test} \label{theme: anxiety}

\begin{figure*}[ht] 
  \centering
  \includegraphics[width=\linewidth]{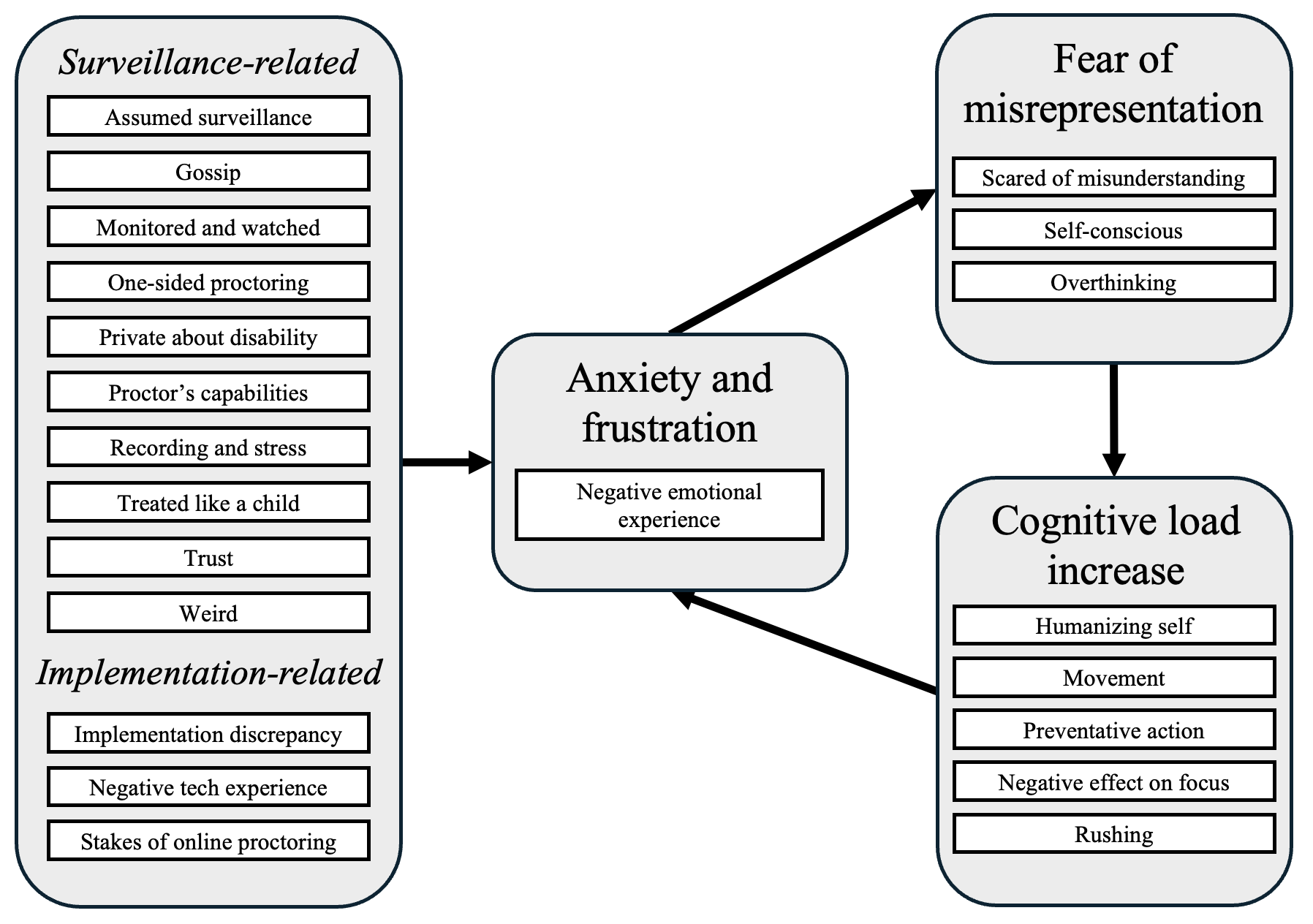}
  \caption{The surveillance and implementation causes of anxiety feed into a cycle of misrepresentation and cognitive load increase.}
  \label{fig: anxiety}
  \Description{The image description can be found in Appendix~\ref{image des: anxiety}.}
\end{figure*}

Surveillance in the OPS creates a situation where students feel anxiety and frustration due to a variety of factors that we coded for. These factors, both explicitly related to surveillance and some to the implementation of the OPS, can be seen in Figure~\ref{fig: anxiety}, which shows the list of codes that contribute to anxiety on the left side. Anxiety impacted the test-taking experiences by increasing fears of misunderstanding, self-consciousness, and overthinking. This would lead to additional behaviors and distractions that increase the cognitive load to take a test, leading to frustration and more anxiety. Figure~\ref{fig: anxiety} maps the codes in our analysis to their place in this model of the impact of surveillance anxiety on testing.

Most students in our study reported having testing anxiety regardless of the OPS surveillance. Mental health and psychological disabilities, like anxiety disorders, were the most reported disability type in our study, as reported in Appendix~\ref{demographics}. Test anxiety was reported during the interviews by twelve participants as seen in Table~\ref{tab: anxiety}.

\begin{table}[ht]
  \caption{Emotion reported during testing in general}
  \label{tab: anxiety}
  \begin{tabular}{cc}
    \toprule
    Emotion & Frequency\\
    \midrule 
    Anxiety & 9 \\
    Stress & 3 \\
    Not asked & 3 \\
    Other emotion & 1 \\
    \end{tabular}
\end{table}

When asked about the emotions they felt toward the OPS, specifically, students reported a variety of emotional experiences that can be summarized as anxiety. P5 and P9 explicitly called the negative emotional experience ``anxiety,'' but it was also described as nervousness (P6) and panic.

\begin{quote}
   ``It just feels like when you're being watched, you can't breathe. You can't catch your breath, racing heart and dry mouth, and then you just panic. And even though you're sitting at your very own desk, you feel like just getting up and running away.'' (P14)
\end{quote}

P4 also described feeling like they ``need to run away'' and P16 also called the emotion ``panic.''

Negative emotional experiences aren't always anxiety, but sometimes they are frustration (P3, P13) or annoyance (P17), which can be caused by technical issues and false flags. P3 explained that she was frustrated because of the unfair environment OPS creates for students with testing anxiety, especially those who have an accommodation for anxiety.

\begin{quote}
    ``The hyper-structured testing environment has exacerbated my test anxiety and made it harder for me to succeed in ways that are unrelated to how well I understand class contents. If the hardest thing about the test is the test environment, and not the content, then we're making an impediment for students to access and participate in their education.'' (P3)
\end{quote}

The experiences of typical test anxiety, anxiety disorders, and the anxiety participants attributed to surveillance are interwoven.

\begin{quote}
    ``The camera being on all the time would, especially if they have anxiety disorders like I do, increase the anxiety levels a little bit. Even though we know we're not doing anything wrong, it just makes us a little bit more nervous. And it affects our accuracy on those assessments and tests.'' (P6)
\end{quote}

\subsubsection{Surveillance-related causes of anxiety} \label{subtheme: privacy-related}

When asked to describe the OPS, many participants used the terms ``monitored'' and ``watched.'' The feelings of being watched and being anxious made up a cycle where students felt uncomfortable being watched in an anxious state while feeling additional anxiety from the monitoring.

\begin{quote}
    ``I'm already doing a thing that is anxiety-inducing and tests are like that for me. So I just don't like that feeling of being watched.'' (P9) 
\end{quote} 

\begin{quote}
    ``I'm not good at multiple choice testing. So I don't know how many people watching me crash and burn, or like get frustrated.'' (P1) 
\end{quote}

It was common for participants in our study to need extra time on tests per their accommodations. Taking a test slowly adds to the stress of being watched because of potential judgment from the online proctor.

\begin{quote}
    ``I don't like being looked at. I don't like feeling under pressure. I like taking my time, and I don't want someone to be like, `Oh, she's taking forever. My gosh! Like so slow,' that sort of thing.'' (P9)
\end{quote}

This could lead students to feel discomfort with the proctor knowing they have a disability. Participants didn't want to be seen as ``dumb'' due to their learning disability (P1) or seen ``in a poor state'' due to their cognitive disability (P16).

Furthermore, being watched left a few participants (P2 and P15) worried about being gossiped about by the online proctors.

\begin{quote}
    ``Someone somewhere on the back end is watching people and making fun of them or laughing at them because of how they look while they're working on a test.'' (P2) 
\end{quote}

P2 worried that her disability-related behaviors, such as chewing on her nails or playing with a fidget toy, would become ``water cooler gossip.'' She also made a connection to being a veteran because she was concerned that a proctor might mock or view as inappropriate the t-shirts she wears that depict ``bloody dog tags'' or references to being a woman veteran.

When asked to define online proctoring, many students highlighted the one-sided nature of the system by describing the proctor as ``someone you never see or talk to'' (P4) and someone you can't ``see what they're doing'' (P8). This is comparable to the panopticon theory of surveillance mentioned in Section~\ref{panopticon}. A reframing of this metaphor can be seen in a couple of participants who described themselves as feeling like they are in a fishbowl. 

\begin{quote}
    ``It's like watching a frog through a fish tank, and seeing whether they're eating or things like that. But you're watching a person take a test.'' (P4)
\end{quote}

\begin{quote} 
    ``It's like I'm in a big fish bowl. There's a reason [...] why I have walls and doors with a lock on it. It's not all glass. I'm not on display with some pop-up art or something. For someone that I don't know to be watching me take an exam just seems wrong.'' (P14)
\end{quote}

Participants reported searching for a way to make contact with the proctor, but failing to find a mechanism to do so due to the one-sided nature of the OPS. P2 told the story of how they attempted to make contact with the proctor to inform the OPS that they needed to go to the bathroom to change their wet diaper. When they were unable to contact the proctor, they finished the exam in the extremely uncomfortable state of sitting in a wet diaper. Other participants described narrating their actions when they moved or stood up from the exam to meet a need, hoping their explanation would be heard by someone.

Not only did participants see themselves as being treated like fish or frogs, but they also described feeling like they were being treated like children. The increased surveillance was ``condescending'' (P3) or in P1's words:

\begin{quote}
    ``You're not a child, and you have the capability to know better than a 6-year-old. You are an adult. The minimum age in [the testing center] is 18. So you understand how this works. You're not new to this. It's like getting your parents to trust you to stay out later. I just prefer not to be slightly micro-managed. [...] I think you should have more responsibility, and it feels like they don't trust us to be responsible for ourselves and actions.'' (P1)
\end{quote}

Similarly, participants described the intensity of the monitoring as demonstrating a lack of trust in them that they felt entitled to as students in HEIs.

\begin{quote}
    ``[It's like] needing a pass to leave the room. In college, we trust that you're an adult and that you can take care of your body and your needs as they come up. That's part of the learning contract that we were making together, and it feels like the opposite of that to sit around and be like, \textit{I have to pee so bad that I can't think about the test right now.}'' (P3)
\end{quote}

Students also considered the OPS as ``weird'' (P1 and P2). P14 said, ``There's just something that's very visceral and gross about surveillance.''

We asked participants what data was shared about them, and the answers were often hesitant because many students didn't know how they were being surveilled. They speculated on the capabilities of the OPS. For example, a notification to close a VPN running in the background indicated that the OPS was monitoring background programs (P11). A blind student at a testing center speculated that they were being video monitored, but didn't know if it was just a webcam or if there were in-person cameras at the testing center, a fact sighted students would more easily ascertain (P9). Many students commented on how the lack of transparency was disturbing. This is discussed more in Section~\ref{theme: values}. The difference between a blind student who has less transparency about surveillance and a student with an anxiety disorder that is hyper-aware of surveillance demonstrates how individual differences of disability have different effects on their perceptions toward surveillance.

\subsubsection{Implementation-related causes of anxiety} \label{subsection: imp}

Some causes of anxiety and frustration were related to the implementation of the OPS. In these instances, OPS was used inconsistently, communicating a lack of consideration for when it is appropriate to implement. Technical issues in the implementation of OPS surveillance were also a trigger for anxiety. Finally, the stakes of OPS implementation, especially in the context of the inconsistent use and unreliable technology, made participants feel even more anxious.    

Participants had a negative outlook on the OPS because they felt it was arbitrarily implemented. 

\begin{quote}
    ``I think it's sort of confusing, and adds to the idea that it feels arbitrary, and having other classes that don't have that system makes it clear that it's optional in some ways for a professor to choose the ways that they want to be strict and enforce academic integrity.'' (P3)
\end{quote}

The technical implementations of OPS caused issues for students that often resulted in anxiety and frustration. Unreliable or slow wi-fi was an issue for some students. Issues with the software itself were hard to resolve. For example, P11 tried to contact their instructor about technical issues with the OPS, but was redirected to contact the OPS company. P11's reaction was: 

\begin{quote}
    ``That's so inappropriate! Just give me a paper exam at this point. At least I could write something.'' (P11) 
\end{quote}

Participants (P11, P13, P17) reported that technical issues would take up time during the testing session, which was especially disruptive to students who were already supposed to be given extra time on exams. Instead of testing, they had that time taken up by trying to resolve technical issues. Technical issues could also be caused by poor integration with accessibility tools, like optical character recognition (OCR), dictation, or magnifying software. The OPS would cause false flags, interruptions, or an inability to use the tools.

Participants (P9 and P14) pointed out that all these anxiety-provoking factors are amplified by how high the stakes are for online proctored exams. Their grades and futures depend on good performance, so an interruption was extremely stressful. P14 explained the self-talk they used to continue the exam after a panic attack.

\begin{quote}
    ``I wasn't even sure if I wanted to continue at that point, but then I thought, \textit{I've gotten all these degrees and I always wanted my Ph.D., so I'm going to finish this no matter what.}'' (P14)
\end{quote}

\subsubsection{Fear of misrepresentation during the test-taking experience} \label{subsection: effect}

Anxiety and negative emotional experiences manifested as concerns about being misunderstood as cheating, self-consciousness about what actions would be seen as suspicious, and overthinking about behaving in a way that wouldn't cause flags. 

\begin{quote}
    ``I was just more concerned if they were gonna flag me for looking away because they give us scrap paper. I didn't want them to misinterpret me as cheating or something.'' (P11)
\end{quote}

Even simple behaviors, such as closing their eyes, could be a trigger for worrying if they were suspicious.

\begin{quote}
    ``And it's like, could I just sit here in my chair and close my eyes and visualize the things I needed in my mind for a while, or is somebody gonna think I'm suspicious?'' (P14)
\end{quote}

Behaviors that are allowed on the exam, like checking outside notes, could also be a trigger for overthinking because of the lack of clarity in the type of surveillance:

\begin{quote}
    ``I always heard that professors can see when you leave the test so I don't want them to be notified that [I am] not on this page for XYZ time. And then I pop back on and it's like my brain was overthinking. But in my head, I'm like, \textit{Oh, that's gonna look weird.} Some [exams] are open notes but not open internet, and I don't know what to do with that information because you do leave your page. But I don't think I worry about that anymore, because I'm like, \textit{It is open note, open book. And my book is on the internet.} So I definitely was just overthinking it.'' (P1)
\end{quote}

Students felt the need to self-regulate and limit their true behaviors because they were anxious about the OPS monitoring. 

\begin{quote}
    ``You have to be cognizant of your body placement and where your eyes are going.'' (P5)
\end{quote}

This was especially problematic when a disability related movement caused flags in the OPS, like when P17, who is in the habit of leaning toward the computer screen to see better, tried to be mindful not to do so during an exam:

\begin{quote}
    ``I tell myself, \textit{Stay calm. Stay still. Don't try to trigger anything needlessly.} But some of my tests are long, so I just get back into the habit.'' (P17)
\end{quote}

This was also a concern for students who stim, make repeated movements, and/or sounds~\cite{cambridge_advanced_learners_dictionary__thesaurus_stimming_2013}. P14, who described having a panic attack triggered by the OPS, explains the link between anxiety and self-consciousness due to the proctoring.

\begin{quote}
    ``[I try] to remain really still and look straight ahead at my screen. It's really difficult to do though when you have ADHD and stimming. It's like I can't. And then when you're already hyped up with anxiety, it makes it worse. My hands are flapping and my feet are [moving]. And then you start thinking, \textit{These people watching me, they probably think I'm crazy or something,} and the mind overacts and then it spirals.'' (P14)
\end{quote}

This quote exemplifies how anxiety and the negative emotional experiences tied to surveillance lead to attempts to regulate behavior. The effort of regulating a behavior causes increased cognitive load on the student. 

\subsubsection{Cognitive load} \label{subsection: cog load}

The intention of using an OPS is to limit and regulate cheating behaviors. However, students suspect that even benign movements could raise alarms even when they are not cheating. This leads students to put extra effort into remaining still during the exam. P6 had previous experiences being flagged for stimming and moving out of the camera frame due to his disability. He described the following feeling:

\begin{quote}
    ``Nervous, cause I didn't want any involuntary movements to cause me to not do well on that test.'' (P6)
\end{quote}

Movements associated with participants using their disability accommodations would also cause flags.

\begin{quote}
    ``I do typically get more flags than other students when I've talked to them after the test if they've had that issue. I would attribute that to me using my accommodations in the moment, and [the OPS] not knowing what's what.'' (P17)
\end{quote}

P17 went on to tell a story about how her test was terminated midway because she hit the limit of how many flags are allowed. P17 attributed this to a more strict live proctor. When P17 tried to explain that the flags were due to her clicking the screen magnifier and moving her body toward the screen to see better, the live proctor told her that they were just following the procedures by flagging her. P17 was able to retake the exam; this time, a more lenient proctor did not flag her as often.

The participants in our study had a handful of actions they took to prevent a false flag of the OPS: laying out their materials (P2 and P6), ensuring their appearance was professional or presentable (P2), staring intently at the screen (P4), sitting up straight (P5), staying in range of the camera (P5 and P6), and breathing techniques (P6).

\begin{quote}
    ``I'm a really bad nail biter, so I put a little sign up on the wall: \textit{Keep your hands out of your mouth.} You just want to look your best or be your best before the camera, not even... not cheat.'' (P2)
\end{quote}

Another strategy was to explain their behavior or humanize themselves to try to influence the OPS not to flag them.

\begin{quote}
     ``I'll get up to get [my pain-relieving pills] and then explain, \textit{Here's the pain reliever. This is what I went down to go get and I'm gonna take it right now. That's all I was doing and if you have any questions, it's in my accommodations. You can talk to the accessibility office.} And usually they're fine with that. Sometimes, [I] also leave a comment through Canvas, which is what our university uses. Or I'll shoot them an email saying, \textit{Hey, this happened during the test.} I explained it just to double triple check to make sure that it's cool.'' (P15)
\end{quote}

One participant used expressing their emotions as a method to make themself relatable to the online proctor.

\begin{quote}
    ``I was under this situation where I was being observed, and I wanted to make sure that they knew that the frustration was real. [...] I wanted to seem almost... humanize myself to them at the same time.'' (P4)
\end{quote}

The participants' efforts to regulate their behaviors were also paired with effects on their focus. Seven participants reported negative effects on their focus because they felt anxious or were focused more on regulating their movement and behavior instead of the exam. One student worked with the student accessibility office to report the difficulty focusing and was able to opt out of using video monitoring on the OPS.

\begin{quote}
    ``I explained to [the disability office advisor] the feeling of how hard it was for me to concentrate on [the test], because I was so focused on somebody watching [me].'' (P8)
\end{quote}

P5 described rushing through the exam to alleviate their uncomfortable feelings about being monitored.

\begin{quote}
    ``I wanting to get it over with ASAP, and that does relate to my anxiety for sure, so I'm anxious about that. I just want to get it over with. [The monitoring] is making me uncomfortable. I want it to be done.'' (P5)
\end{quote}

Another reason for rushing was described by P17, who felt that if they took the test fast, there would be fewer flags and they might get more favorable outcomes from the OPS and their instructor reviewing the flags.

\begin{quote}
    ``I just try to make it so when my instructors review it they aren't reviewing multiple hours of footage. That way there's overall less flags, and this is a shorter file or instance they have to review. Just making it easier on them.'' (P17)
\end{quote}

\subsection{Conflicting perceptions about the surveillance-privacy tradeoffs that students with disabilities experience with OPS} \label{theme: compromise}

\begin{figure}[ht]
  \centering
  \includegraphics[width=\linewidth]{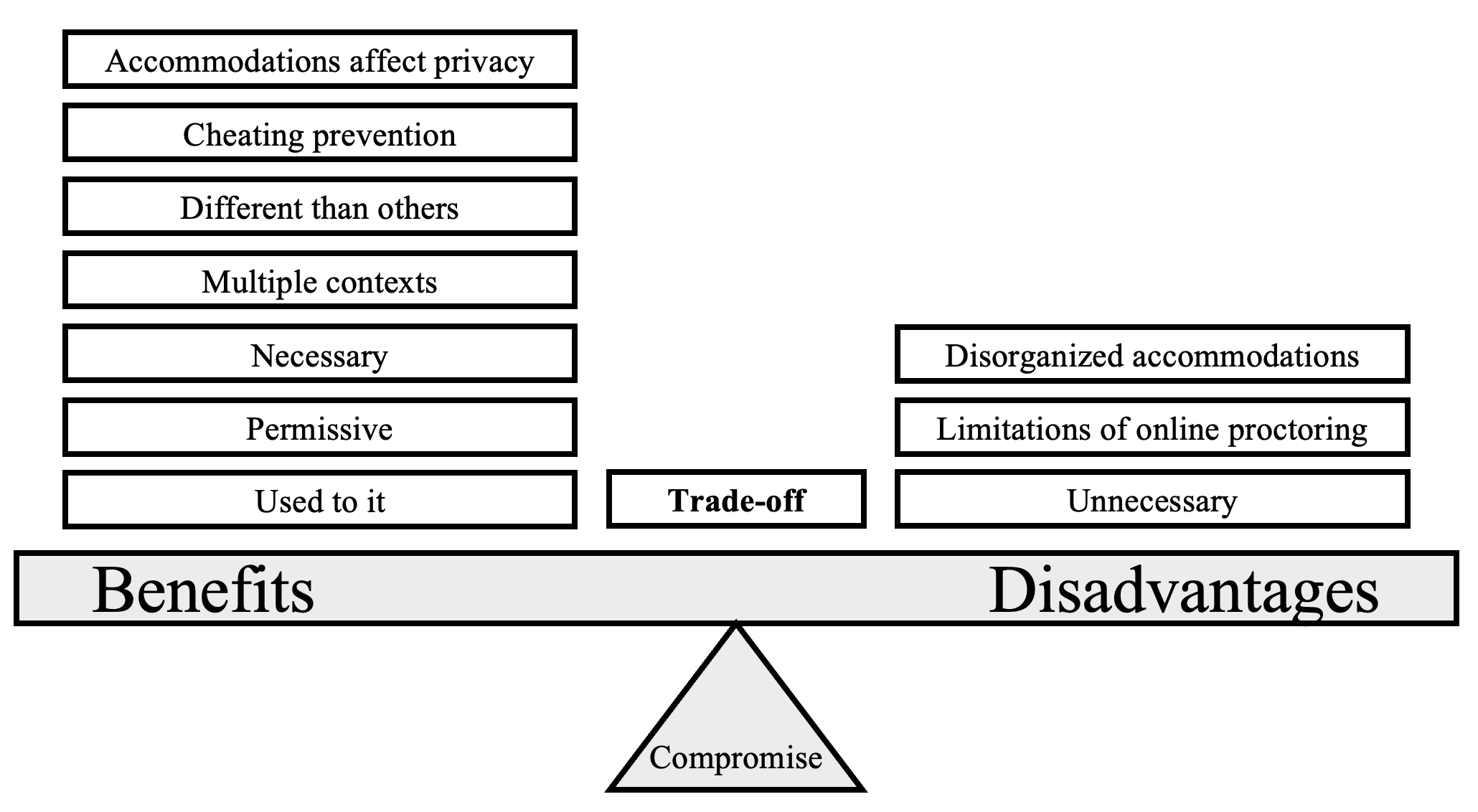}
  \caption{Participants balance benefits and disadvantages in their perceptions of OPS.}
  \label{fig: compromise}
  \Description{The image description can be found in Appendix~\ref{image des: compromise}.}
\end{figure}

While most students reported negative emotional experiences or a dislike for surveillance, students also weighed the benefits and disadvantages of using OPS to take tests. Although students don't have a meaningful choice whether to use an OPS, they weigh the positives and negatives to generally accept its use. The codes in this theme are illustrated in Figure~\ref{fig: compromise}.

\subsubsection{Benefits of OPS use} \label{subtheme: pro}

Some students felt an increased level of privacy because of their accommodations. For example, most students need to download the OPS onto their personal computers, but P17, who goes to a testing center to use the OPS, could use the computers at the testing center. This alleviated privacy concerns about having intrusive software on a personal computer. P1 also pointed out that going to a testing center and getting her grades back electronically made it so she wasn't pressured to share her grades with other students.

Most participants brought up that they are willing to make privacy sacrifices, or understand why they are being asked to make privacy sacrifices, for the goal of cheating prevention. 

\begin{quote}
    ``I don't really enjoy it. But I also understand why they do it and why it's necessary due to cheating.'' (P13)
\end{quote}

However, participants felt the OPS was unfair because they were flagged more compared to their non-disabled peers, so ideally, they would have more lenient forms of proctoring.

\begin{quote}
    ``I understand some of my instructors need [to use OPS]. Especially with the rise of using AI to cheat, in my major, I've had students openly tell, \textit{I'm using ChatGPT to write this.} And I'm like, \textit{Okay...} So I understand the necessity for it. I just wish, and I know that's asking a lot because these instructors have all these students, I'm not that significant. But I know I'm not doing that. I know that I'm just using my accommodations. It'd be easier if I could do it at home. But in general, I know that a lot of these external factors cause that.'' (P17)
\end{quote}

Other students brought up that they saw themselves as different than other students who might have more privacy concerns (P11) or less knowledge about privacy (P13).

\begin{quote}
    ``Thankfully for me, I don't have much things to hide. But for somebody who may feel a lot more sentimental with their stuff, they probably would not want a software to know something not related to school.'' (P11)
\end{quote}

P13 felt exceptional in his knowledge about privacy. He worked with the information technology staff at his HEI to ensure he was accommodated with the least possible privacy issues.

\begin{quote}
    ``I guess to strike the balance of security and accommodation, I think it's pretty good. I don't see how they could make it really any better because they'd have to somehow integrate it into an existing browser that would not allow it, at least for my particular needs. And I also acknowledge that most people do not care about having an extra browser on their computer. The other students that I've seen, they have like six hundred apps on their phone, six hundred applications installed on their computer. They don't care, so I think it's more so a particularity thing for me. I think they do about as good as they can.'' (P13)
\end{quote}

P12 considered OPS monitoring (done from home) the same as monitoring in a school context, which was okay with them in the same way they would be okay with monitoring at police stations and hospitals. Similarly, P8 thought that exam monitoring was similar to monitoring at work, where an employer needs to see what they are doing.  

\begin{quote}
    ``Screen recording is not something that you use all the time. It's only for jobs or school, not for just for anybody to watch you over.'' (P8)
\end{quote}

Participants who considered themselves more permissive with the proctoring had a few reasons. 

Although in Section~\ref{subtheme: privacy-related} we described students nervous about one-sided proctoring, P11 thought that the way the OPS was designed not to see the proctor was a positive.

\begin{quote}
    ``What's really good about it is most of these applications do not show the proctor staring at you. If it was the case, that would be a bit more of an issue. I think it's a lot less invasive, in my opinion.'' (P11)
\end{quote}

P11 and P16 both felt that they could forget about monitoring. P12 felt permissive because, unlike many other participants, he felt that he had been given the chance to consent to the monitoring.

\begin{quote}
    ``I consented to it as part of the deal. To me, that'd be like me going back on the deal. It's like a mental pep talk. Even if [I] didn’t like it, which I don't care, it's like, \textit{Well, this is how it's going to be, so suck it up. Deal with it.}'' (P12)
\end{quote}

Another participant (P2) talked about how she is used to monitoring due to her disability, which makes her used to monitoring in testing contexts too. 

\begin{quote}
    ``If it's not about cheating, it's about, \textit{Is she going to fall? Is she going to get hurt? Is she safe?} It's just that constant monitoring, \textit{Is she okay?} That kind of stuff that you overall get used to being watched over anyways, and so it felt like an extension of that.'' (P2)
\end{quote}

P1 and P6 also used the term ``used to it.''

Participants saw OPS as a reality of modern-day testing, so they understood it to be a necessity.

\begin{quote}
    ``It’s like everyone's doing online proctoring now, so it's kind of like you don't have a choice but to do it.'' (P5)
\end{quote}

\begin{quote}
    ``Any sort of frustration, it doesn't come out towards that because I understand the necessity of it.'' (P12)
\end{quote}

Despite a range of emotions toward surveillance, some students thought that OPS were necessary due to cheating prevention, appropriateness for the context of education, and the normalization of surveillance. Other students felt generally permissive, used to the monitoring, different from other students in their concerns, or even more private using an OPS in a testing center. These factors can be seen on the Advantages side of the scale pictured in Figure~\ref{fig: compromise}.

\subsubsection{Disadvantages of OPS use}

On the contrary, some students considered the OPS unnecessary because it did not reflect the realities of work outside of school.

\begin{quote}
    ``The real world has resources, and so also, as a student, you should have resources. I also think that part of the goal of education and academic integrity is to instill intrinsic academic integrity, not because I'm afraid I'm going to get caught, but because I'm confident in myself and my ability to express my own ideas, and because I care about being here and doing the best job that I can rather than cheating.'' (P3)
\end{quote}

There was confusion about the implementation of OPS when it didn't seem necessary.

\begin{quote}
    ``Sometimes when we have tests that are open note, open book. I don't really see the point in being proctored because I have everything available to me.'' (P1)
\end{quote}

Participants additionally felt that the OPS was unnecessary because it interfered with their accommodations by cutting off their allotted time (P1), taking a long time to implement their accommodations (P3), not allowing them to implement their accommodations at all (P4), or requiring them to request their accommodations every time they took an exam (P17).

P5 was against the implementation of OPS because he found the limited scope of the OPS problematic: it doesn't show the ``full picture'' to know if someone is actually cheating.

The OPS being unnecessary, causing interruptions to students' accommodations, and it's limited scope were the disadvantages of using OPS for our participants

\subsubsection{Trade-off between surveillance and benefits}

Overall, many students saw benefits and disadvantages to using OPS, so they viewed OPS as a trade-off between these positive and negative factors. Figure~\ref{fig: compromise} reflects that the trade-off code sits in the balance of compromises that need to be made between OPS surveillance, implementation, and accommodations. 

For example, despite preferring in-person, paper exams, P1 thought using an OPS was fine for take-home exams. P6 also preferred in-person exams and saw the privacy issues as a trade-off for taking an exam at home.

\begin{quote}
     ``I guess there's a trade-off from being able to do it with the professor. Because you're kind of giving away that personal data just to access your exam. And you're not really sure if the company has been subject to any data breaches or whatever. So you're kind of taking a risk.'' (P6)
\end{quote}

A similar sentiment was echoed by P8.

\begin{quote}
    ``I understand that certain stuff is not private. And I understand that the reason for this is that if they're giving us the opportunity to be able to work from home.'' (P8)
\end{quote}

P5 liked the convenience of being able to take online exams any time he wanted, but pointed out that he would procrastinate taking an exam if he had the choice of when to take it. Choosing when to take the exam can be beneficial to students with chronic illnesses who need more flexibility.

\begin{quote}
    ``I think it does open up the opportunity for individuals that need to test differently that we didn't have before. So I think that is going to be very beneficial for those type of individuals, especially online. Online offers the ability for people like, if I'm sick and I have bronchitis and stuff and can't go in, or even when I'm in the hospital, I took one of my quizzes when I was in the hospital. You know, it offers that opportunity for remote learning in different ways, whereas before we would have to miss or drop out of.. because you miss so much, etc. This opens the doors for more people with varying situations to be able to participate in the academic process that wasn't open before. So I think it's been real, beneficial.'' (P2)
\end{quote}

Despite participants reporting that their accommodations were not always properly implemented, OPS has limitations that may make it ineffective, and OPS is not necessary due to unrealistic implementation in real-world situations. Participants had a multitude of reasons to accept the use of OPS. They would sometimes feel more private on the OPS or in the testing center. They wanted there to be cheating prevention. They saw themselves as different in their privacy concerns from more affected students or as more generally permissive people. They were used to the testing on OPS. They also saw OPS as appropriate to the school context, and that it was necessary due to it being the current paradigm for test-taking. Overall, students reported understanding the trade-offs of more surveillance for flexible test-taking. The balance was different for every participant.

\subsection{OPS violate students' privacy values and share information they would rather keep private.} \label{theme: values}

\begin{figure}[ht] 
  \centering
  \includegraphics[width=\linewidth]{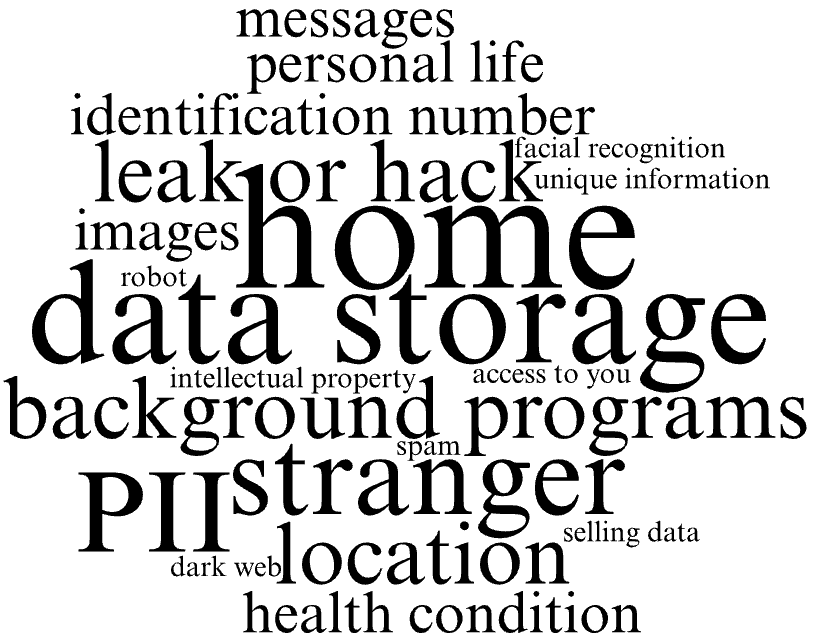}
  \caption{The prominence of various topics brought up as privacy vulnerabilities.}
  \label{fig: wordcloud}
  \Description{The image description can be found in Appendix~\ref{image des: word cloud}.}
\end{figure}

The values of control, protecting others, avoiding societal targeting, transparency, and consent were prominent in participants' descriptions of how they defined privacy and what privacy protections they desired from the OPS. A few students defined themselves as ``private people'' but privacy vulnerabilities came up in almost every interview. A compilation of those vulnerabilities resulted in a large number of references to what kind of information students valued keeping private. When asked about their definitions of privacy, participants gave a multitude of factors, as seen in Figure~\ref{fig: wordcloud}, with privacy about their home, the storage of their data, their PII, and issues with the OPS proctor being a stranger as the most prominent. 

\subsubsection{Values}

Students wanted a choice in what features of the OPS they were using, like control over their computer and mouse (P4, P9), collecting background data (P13), their data being used to train future iterations of the OPS (P15), and having their camera on (P16). P4, P14, and P16 brought up agency, autonomy, and freedom, which, in their experiences of disability, are especially important values. 

\begin{quote}
    ``Whenever I don't feel I have a choice, I start to feel really, really bad and I want to take back control because I don't like the idea of anybody controlling my own agency or controlling me in any way. Freedom is very central to me because of what I dealt with. I value freedom more than anybody else realize.'' (P16)
\end{quote}
 
P4 also pointed out the intersection of disability and race by comparing proctoring to policing and how police target people with disabilities and people of color.

\begin{quote}
    ``The way that I mentally connect policing and proctoring [is that] they're in the same vein to me, in terms of being watched, in terms of having someone over your shoulder, in terms of being different to other students or potentially singled out. My experience with this comes from being in environments where I was one of the only Southeast Asian people there, in a predominantly white room, and then, being very concerned as a student, as a kid who didn't know much, that I was already different. I could feel being different all these different ways, and I could feel that people were treating me with a tenseness and a sense of difference and a sense of tolerance, but not acceptance, that extended towards proctoring.'' (P4)
\end{quote}

On the other hand, another participant (P11) was worried Asian Americans would be given higher scores if the collection of race information by the OPS was part of the grading process. 

Alongside choice and control, students valued consent to use the OPS. 

\begin{quote}
    ``Your ability to keep the information you want to yourself and share the information that you want to feels very tied to consent, and what you personally agree is information you're willing to be shared about yourself and not sharing the information that you're not willing to be shared.'' (P3)
\end{quote}

A complicating factor for consent was the lack of transparency, making it difficult to understand what information was being collected. P6 said their professor mentioned what data was collected by the OPS at the start of class, others said privacy information was probably in the terms and conditions of using the OPS (P5, P1). Regardless, these students and others asked for more transparency. For example, participants said they were told there was recording or video monitoring, but it was unclear what kind.

\begin{quote}
    ``They didn't say whether they were recording people. They just said that the cameras would be on, and people would be watching, but they didn't say who. They didn't say if they were recording or saving any of this, so that was a mystery.'' (P14)
\end{quote}

\begin{quote}
    ``If I was told, \textit{It's a recording}, but then I find out later somebody was watching that whole time. Even though one could argue, \textit{You're supposed to just be taking the test. What does it matter?} I like to be aware if somebody's watching versus it being recorded as a general thing because I don't know why that person's watching. When it is disclosed that someone's watching, it clearly says, for the purpose of taking the test and monitoring the test taker. If somebody's just watching, and I had no idea, it's like you could be watching for a different purpose, and I wouldn't know because there wasn't any disclosure.'' (P17)
\end{quote}

Transparency was especially salient for blind participants who had even less information about if they were being monitored by surveillance cameras in the testing center (P9) or if there was a live proctor shown on their computer screen.

\begin{quote}
    ``Since I can't see, I don't think you could see the proctor's face either, but since I can't see someone's face, I'd like to hear them speak, or background noise to know they're there, or else it's just nothingness, so some signal, some sign, that there's an actual person there, a normal person who doesn't have bad intentions or something like that.'' (P9)
\end{quote}

Another value held by the participants in our study was protecting children, or other people who were not the intended targets of monitoring, from the OPS surveillance.

\begin{quote}
    ``I wouldn't mind them seeing [my room], there's nothing bad about it. It's all clean and normal. But I just wouldn't want my kids or friends unwillingly on camera.'' (P12)
\end{quote}

\begin{quote}
    ``[I want to] make sure other people's business isn't getting put out there. [...] I trust the [school] pretty well, it's not [...] that [my information is] just going to a site on the Internet I would not trust that at all. So the school, it's a little more trustworthy, but I still don't want my kids to think they don't have their own privacy while I'm doing [the exam].'' (P8)
\end{quote}

\subsubsection{Vulnerabilities}

A few participants described themselves explicitly as valuing privacy due to being a private, privacy-minded person (P9, P13), liking to be anonymous (P16), or being raised in a generation where surveillance was not as commonplace (P12). Regardless of whether they saw themselves as privacy-minded, almost every participant (n=14 out of 16) told us what information they would find vulnerable to share. The most popular concern was information about their home, room, or personal space in the background of video monitoring.

\begin{quote}
    ``I think it's sometimes weird because I've also seen videos of people like you have to prove there's nothing in your room and show around your room; and I'm like, \textit{That seems so weird and uncomfortable.} I don't need my professor like that... This is my room like.. Yeah, that's definitely private and personal.'' (P1)
\end{quote}

Students were also concerned about how their data was being stored.

\begin{quote}
    ``What would happen with video audio? What happens with that? Are they recording it? Are they saving it? How long are they saving it? Who views it? Who else gets to view it? Who has access to this? It's about privacy of my personal data.'' (P14)
\end{quote}

Students were also concerned about their data being leaked or hacked.

\begin{quote}
    ``How long do these recordings get stored? I'm assuming they're being taken and stored on someone's server somewhere so that they're available, so that, if the professor wants to go back and watch everyone take their test, they can. I have no idea how long they store that for and especially with all the AI training now, who's to say if that just gets leaked or someone steals that data and they just have hours and hours of footage of me just taking tests. Not a big fan of that.'' (P15)
\end{quote}

Participants wanted personal identifying information (PII) to be kept secure, especially if it was related to their disability.

\begin{quote}
    ``It's your identifying information. I would say things that are related to you and things that are generated by you, like your date of birth and everything like that, that type of information, but also includes your actions and your thoughts and behaviors. That's your privacy, I feel like anything that relates to you and that's unique to you. [...] It's easy for them to observe I'm a girl. They can see I'm a girl, fine. But let's say it's more nuanced like, they don't know my history or like my health condition.'' (P11)
\end{quote}

The fact that the OPS live proctor was a stranger was also uncomfortable for participants. This interacted with the discomfort of showing their home, room, or personal space.

\begin{quote}
    ``Because this is my own private space. I mean, it's some stranger that I don't know, that I don't see them, is watching in my own personal home while I take an exam, and it just feels wrong.'' (P14)
\end{quote}

The negative effect of having a stranger watching was the lack of personal connection, which could mean more flags or bad intentions

\begin{quote}
    ``Well, you're being watched, but it's like super supervision, like increased supervision more than I would have if I was in class with a professor, only because I can't defend myself. Because the person who could report me is a complete stranger who has no allegiance to me. So yeah, it's definitely a negative. It has a negative effect.'' (P5)
\end{quote}

The intentions of the OPS proctor are unknown because the proctor is unknown.

\begin{quote}
    ``I feel more comfortable with the professor because I know everybody in the classroom. Whereas [using the OPS at the testing center], I have no clue who's proctoring and watching over you. It could be absolutely anywhere. And even now, in the day where things are so politically, you know. Can I say this and not have it come back up again and show up in a way that it could hurt me?'' (P2)
\end{quote}

The other types of information students felt vulnerable about sharing were their location, the programs running in the background of their computer or other open tabs, messages, their images, their social security or driver's license numbers, access to unique information about them or ways of contacting them, their intellectual property such as research reports, facial recognition, and non-school related information. They were worried about this information ending up on the dark web, being sold, or used to train AI OPS. A summary of these issues can be found in Figure \ref{fig: wordcloud}.

\section{Discussion and Implications}

Although every difficult experience with OPS shared with us by our participants is uniquely worthy of being remediated, there are a few design implications for OPS that could benefit many students with and without disabilities.

\subsection{Use contextual integrity to address asymmetric information sharing}

Other education technology, such as apps~\cite{hasan_understanding_2023,khan_exploring_2023}, learning management systems~\cite{kwapisz_privacy_2024, jiang_data_2023}, and learning analytics~\cite{slade_learning_2019,prinsloo_student_2015}, have been described as creating power asymmetries between students and instructors and information asymmetries between the students and the education technology companies that gather, store, and use data. Power is core to understanding the results of this study because of the framing of surveillance through Foucault's philosophy~\cite{foucault_powerknowledge_1980}. The prisoner in the panopticon, not knowing if they are being watched, is as disturbing as being constantly watched because of the power the watcher has over them~\cite{foucault_discipline_1977}. This is echoed in our data by the one-sided relationship with the proctor, like in the quotes in Section~\ref{subtheme: privacy-related} about being in a fish tank. Students being flagged for cheating or not performing well on an exam has a large impact on how they perceive the power the OPS has over their grades and futures, as seen in Section~\ref{subsection: imp}. This is echoed by Balash et al. and Selwyn et al.~\cite{balash_examining_2021,selwyn_necessary_2023}.

Power asymmetries are especially important to address for disabled students because of the prevalence of ableism in academia, which puts disabled students in a less powerful position than their non-disabled peers. Disabled students are not likely to have disabled professors~\cite{brown_ableism_2018}. They are put in vulnerable and oppressive positions to get the accommodations they need~\cite{reutlinger_its_2022,roslin_vitriolic_2021,grimes_university_2019}. They are subject to microaggressions that have negative impacts on their academic success~\cite{lett_impact_2020}. The gap in power between students with disability accommodations and the OPS they use is evident throughout their experiences of unfair flags, fear, and violations of their values. 

How do we design to remediate the power and information imbalances shown in this analysis and other academic literature? Starting from the standpoint of privacy as contextual integrity aids in the understanding of how information flows in the OPS and larger stakeholder system~\cite{mutimukwe_privacy_2023}. The expectations and benefits seen by participants in our study are reflected in Section~\ref{subtheme: pro}, where students are willing to sacrifice some privacy in the context of education, expectations, and cheating prevention. Better communication through a privacy nutrition label~\cite{kelley_nutrition_2009} or indicators of what kind of surveillance is taking place are potential interventions. For example, a red light on a camera is an indicator that it is recording; similar signals are needed for OPS. Importantly, these signals need to be accessible to avoid situations like P9's, where she did not know what visual recording was happening because she couldn't see cameras on the ceiling or a video feed on her screen (Section~\ref{subtheme: privacy-related}). Additionally, the implementation of the OPS needs to be appropriate for the testing context~\cite{fawns_matter_2022}. Stakeholders seeking to implement OPS should be mindful of whether the test-taking goals and conditions require an OPS. Once using an OPS is justified by the goals and conditions of the exam, and students have been informed about the surveillance implications, there should be opportunities for students to consent to the use of an OPS. Other researchers have advocated for asking students for permission because it would uphold the transmission principle of contextual integrity~\cite{terpstra_online_2023}. The values of transparency, appropriateness, and consent can all be built into OPS through upholding contextual integrity principles for a flow of information that reduces asymmetries.

\subsection{Address anxiety with ability-based design}

The prevalence of anxiety in our study from both privacy and non-privacy related triggers is consistent with other studies that show increased anxiety on online proctored exams~\cite{conijn_fear_2022,butler-henderson_systematic_2020,kuleva_exploring_2024,james_tertiary_2016,ul_haque_nuanced_2023,pokorny_out_2023}. Privacy and surveillance are shown to be triggers~\cite{conijn_fear_2022,kuleva_exploring_2024} as well as technical challenges~\cite{james_tertiary_2016,ul_haque_nuanced_2023}. Other research shows that students feel anxiety because they feel their privacy is invaded by biometric technology, constant monitoring, and a lack of familiarity with the OPS~\cite{ul_haque_nuanced_2023}, which is consistent with the experiences of students with disability accommodations.

While many studies have investigated the role of anxiety in OPS, few have made the connection to anxiety disorders~\cite{gin_covid-19_2021,kuleva_exploring_2024}. Our findings mirrored Gin et al.'s, where a student described a concept similar to the cycle of anxiety being proctored in a stressed-out state, increasing stress, as seen in Section~\ref{theme: anxiety}. The accommodations to alleviate anxiety were nullified by the increased stress of being on camera. Common behaviors associated with anxiety coincide with the types of behaviors OPS detects to indicate cheating, such as eye gaze shifting, moving the head, and shifting in the seat~\cite{kolski_examining_2018}, which creates an equity concern for students who are more prone to anxiety. Lett et al. found that poorer academic performance among students with disabilities and greater symptoms of anxiety are predicted by experiences of ableism~\cite{lett_impact_2020}.

Ability-based design advocates for creating systems that cater to the strengths of users with disabilities rather than creating adaptations for inaccessible systems~\cite{wobbrock_ability-based_2018}. Similar frameworks~\cite{antonenko_framework_2017} have been used to design for the strengths of autistic users with anxiety in virtual reality systems~\cite{schmidt_toward_2024}. Building on Schmidt et al.'s suggestion, we recommend adding calming routines and meditation into OPS. The participant quoted in Section~\ref{subsection: effect}, who was worried that closing their eyes to visualize would be seen as suspicious, would be encouraged to do so instead of being punished. Meditation has been shown to reduce test anxiety~\cite{lothes_sitting_2023}. More research is needed into how integrating meditation and calming routines into OPS would affect test anxiety, especially for students with accommodations for anxiety.

\subsection{Start from universal design in higher education principles and privacy by design}

Universal design in Higher education (UDHE) combines the principles of Universal Design (UD), Web Content Accessibility Guidelines (WCAG), and Universal Design for Learning (UDL) to advocate for designing educational products for a spectrum of abilities from the start of the design process instead of retrofitting inaccessible technology~\cite{burgstahler_what_2021}. In our discussion of changes in behavior due to anxiety about being perceived as cheating or suspicious (Section~\ref{subsection: cog load}), a few principles of UDHE are violated. Students are flagged more frequently for using their accommodations, and they need to explain their behavior because there is no way to communicate their needs. These are both design oversights because the OPS was not built to integrate with accessibility tools, which violates the principle of operability. The engagement principle is violated when students can not access the proctor to be able to proceed in the exam with their needs met. UDHE is critical for addressing accessibility concerns.

UDHE should be combined with Privacy by Design (PbD) because both frameworks are based on being proactive instead of reactive in the design process~\cite{cavoukian_privacy_2010}. Participants in our study described clear violations of the principles of visibility and transparency~\cite{cavoukian_privacy_2010} when they were not told what kind of data was being collected about them, as seen in Section~\ref{theme: values}. Also reflective of Section~\ref{theme: values}, other researchers have highlighted students' prominent concerns about data storage and breaches~\cite{shioji_its_2025}. Protection of students' privacy values should be integrated into the OPS design process from the beginning.

\section{Conclusion}

OPS violates students’ privacy by implementing overly intrusive surveillance that students cannot meaningfully consent to. The technologies used in OPS are shown to unfairly flag students with disabilities. Our reflexive thematic analysis of interviews with students who take online invigilated exams and have disability accommodations points to their anxiety about the interaction between surveillance and their disabilities, leading to negative changes in behavior and mindset toward the exam. Students express the compromises they need to make with their privacy and accommodations to take remote tests. We present the implications for the design of OPS to mitigate the issues faced by disabled students. We suggest using contextual integrity to address asymmetric information sharing, addressing anxiety with ability-based design, and starting the OPS design process from universal design in higher education principles and privacy by design. Future work is needed on the design solutions to the issues at the center of disability and privacy in OPS, which we present in this article.

\begin{acks}
\end{acks}

\bibliographystyle{ACM-Reference-Format}
\bibliography{references}

@inproceedings{li_visual_2021,
	address = {Yokohama Japan},
	title = {A {Visual} {Analytics} {Approach} to {Facilitate} the {Proctoring} of {Online} {Exams}},
	isbn = {978-1-4503-8096-6},
	url = {https://dl.acm.org/doi/10.1145/3411764.3445294},
	doi = {10.1145/3411764.3445294},
	language = {en},
	urldate = {2025-09-10},
	booktitle = {Proceedings of the 2021 {CHI} {Conference} on {Human} {Factors} in {Computing} {Systems}},
	publisher = {ACM},
	author = {Li, Haotian and Xu, Min and Wang, Yong and Wei, Huan and Qu, Huamin},
	month = may,
	year = {2021},
	pages = {1--17},
}

@article{hernandez_influence_2021,
	title = {The {Influence} of {Test} {Anxiety} and {Stress} on {Gastrointestinal} {Symptoms} in {College} {Students}},
	volume = {116},
	number = {S526},
	journal = {The American Journal of Gastroenterology},
	author = {Hernandez, Oscar L. and Gaisinskaya, Polina and Okpokpo, Eno-Emem and Goyal,, Suhani and Sule, Sachin},
	year = {2021},
	pages = {S526},
}

@article{lin_looking_2023,
	title = {Looking at the {Body}: {Automatic} {Analysis} of {Body} {Gestures} and {Self}-{Adaptors} in {Psychological} {Distress}},
	volume = {14},
	copyright = {https://ieeexplore.ieee.org/Xplorehelp/downloads/license-information/IEEE.html},
	issn = {1949-3045, 2371-9850},
	shorttitle = {Looking at the {Body}},
	url = {https://ieeexplore.ieee.org/document/9506822/},
	doi = {10.1109/TAFFC.2021.3101698},
	language = {en},
	number = {2},
	urldate = {2025-09-09},
	journal = {IEEE Transactions on Affective Computing},
	author = {Lin, Weizhe and Orton, Indigo and Li, Qingbiao and Pavarini, Gabriela and Mahmoud, Marwa},
	month = apr,
	year = {2023},
	pages = {1175--1187},
}

@article{kapp_people_2019,
	title = {‘{People} should be allowed to do what they like’: {Autistic} adults’ views and experiences of stimming},
	volume = {23},
	issn = {1362-3613, 1461-7005},
	shorttitle = {‘{People} should be allowed to do what they like’},
	url = {https://journals.sagepub.com/doi/10.1177/1362361319829628},
	doi = {10.1177/1362361319829628},
	language = {en},
	number = {7},
	urldate = {2025-09-09},
	journal = {Autism},
	author = {Kapp, Steven K and Steward, Robyn and Crane, Laura and Elliott, Daisy and Elphick, Chris and Pellicano, Elizabeth and Russell, Ginny},
	month = oct,
	year = {2019},
	pages = {1782--1792},
}

@inproceedings{balash_examining_2021,
	address = {Virtual Event},
	title = {Examining the {Examiners}: {Students}' {Privacy} and {Security} {Perceptions} of {Online} {Proctoring} {Services}},
	language = {en},
	booktitle = {Proceedings of the {Seventeenth} {Symposium} on {Usable} {Privacy} and {Security}},
	publisher = {USENIX Association},
	author = {Balash, David G and Kim, Dongkun and Shaibekova, Darika and Fainchtein, Rahel A and Sherr, Micah and Aviv, Adam J},
	month = aug,
	year = {2021},
	pages = {633--652},
}

@inproceedings{armagno_role_2012,
	address = {Birmingham, United Kingdom},
	title = {The {Role} of {HCI} in the {Construction} of {Disability}},
	copyright = {http://creativecommons.org/licenses/by/4.0/},
	url = {https://scienceopen.com/hosted-document?doi=10.14236/ewic/HCI2012.71},
	doi = {10.14236/ewic/HCI2012.71},
	language = {en},
	urldate = {2025-09-07},
	booktitle = {{HCI} {Research} in {Sensitive} {Contexts}: {Ethical} {Considerations}},
	publisher = {BISL},
	author = {Armagno, Gustavo},
	month = sep,
	year = {2012},
	pages = {1--4},
}

@article{hasan_understanding_2023,
	title = {Understanding {EdTech}'s {Privacy} and {Security} {Issues}: {Understanding} the {Perception} and {Awareness} of {Education} {Technologies}' {Privacy} and {Security} {Issues}},
	volume = {2023},
	issn = {2299-0984},
	url = {https://petsymposium.org/popets/2023/popets-2023-0110.php},
	doi = {10.56553/popets-2023-0110},
	number = {4},
	journal = {Proceedings on Privacy Enhancing Technologies},
	author = {Hasan, Rakibul},
	month = oct,
	year = {2023},
	note = {2023b},
	pages = {269--286},
}

@article{webber_limitations_2019,
	title = {Limitations in {Data} {Analytics}: {Considerations}  {Related} to {Ethics}, {Security}, and {Possible}  {Misrepresentation} in {Data} {Reports} and  {Visualizations}},
	volume = {2019},
	url = {https://ihe.uga.edu/rps/2019_003},
	number = {003},
	journal = {IHE Research Projects Series},
	author = {Webber, Karen L and Morn, Jillian},
	year = {2019},
	pages = {1--19},
}

@article{depew_mediating_2009,
	title = {Mediating {Power}: {Distance} {Learning} {Interfaces}, {Classroom} {Epistemology}, and the {Gaze}},
	volume = {26},
	copyright = {https://www.elsevier.com/tdm/userlicense/1.0/},
	issn = {87554615},
	shorttitle = {Mediating {Power}},
	url = {https://linkinghub.elsevier.com/retrieve/pii/S8755461509000371},
	doi = {10.1016/j.compcom.2009.05.002},
	language = {en},
	number = {3},
	urldate = {2025-09-07},
	journal = {Computers and Composition},
	author = {DePew, Kevin Eric and Lettner-Rust, Heather},
	month = sep,
	year = {2009},
	pages = {174--189},
}

@book{singer_neurodiversity_2017,
	address = {Lexington, Kentucky},
	title = {{NeuroDiversity} -{The} {Birth} of an {Idea}},
	language = {en},
	author = {Singer, Judy},
	year = {2017},
}

@article{botha_neurodiversity_2024,
	title = {The neurodiversity concept was developed collectively: {An} overdue correction on the origins of neurodiversity theory},
	volume = {28},
	issn = {1362-3613, 1461-7005},
	shorttitle = {The neurodiversity concept was developed collectively},
	url = {https://journals.sagepub.com/doi/10.1177/13623613241237871},
	doi = {10.1177/13623613241237871},
	language = {en},
	number = {6},
	urldate = {2025-09-07},
	journal = {Autism},
	author = {Botha, Monique and Chapman, Robert and Giwa Onaiwu, Morénike and Kapp, Steven K and Stannard Ashley, Abs and Walker, Nick},
	month = jun,
	year = {2024},
	pages = {1591--1594},
}

@article{melian_getting_2022,
	title = {Getting ahead in the online university: {Disclosure} experiences of students with apparent and hidden disabilities},
	volume = {114},
	issn = {08830355},
	shorttitle = {Getting ahead in the online university},
	url = {https://linkinghub.elsevier.com/retrieve/pii/S0883035522000696},
	doi = {10.1016/j.ijer.2022.101991},
	language = {en},
	urldate = {2025-09-07},
	journal = {International Journal of Educational Research},
	author = {Melián, Efrem and Meneses, Julio},
	year = {2022},
	pages = {101991},
}

@article{edwards_academic_2022,
	title = {Academic accommodations for university students living with disability and the potential of universal design to address their needs},
	volume = {84},
	issn = {0018-1560, 1573-174X},
	url = {https://link.springer.com/10.1007/s10734-021-00800-w},
	doi = {10.1007/s10734-021-00800-w},
	language = {en},
	number = {4},
	urldate = {2025-09-07},
	journal = {Higher Education},
	author = {Edwards, Miriam and Poed, Shiralee and Al-Nawab, Hadeel and Penna, Olivia},
	month = oct,
	year = {2022},
	pages = {779--799},
}

@book{oliver_politics_1990,
	address = {London},
	edition = {1},
	series = {Critical {Texts} in {Social} {Work} and the {Welfare} {State} ({CTSWWS})},
	title = {The {Politics} of {Disablement}},
	isbn = {978-1-349-20895-1},
	url = {https://doi.org/10.1007/978-1-349-20895-1},
	publisher = {Red Globe Press},
	author = {Oliver, Michael},
	year = {1990},
}

@incollection{shakespeare_social_2006,
	title = {The social model of disability},
	booktitle = {The disability studies reader},
	publisher = {Routledge},
	author = {Shakespeare, Tom},
	year = {2006},
	pages = {16--24},
}

@inproceedings{shioji_its_2025,
	address = {San Francisco, CA, USA},
	title = {“{It}'s been {Lovely} {Watching} you”: {Institutional} {Decision}-{Making} on {Online} {Proctoring} {Software}},
	copyright = {https://doi.org/10.15223/policy-029},
	isbn = {979-8-3315-2236-0},
	shorttitle = {“{It}'s been {Lovely} {Watching} you”},
	url = {https://ieeexplore.ieee.org/document/11023510/},
	doi = {10.1109/SP61157.2025.00018},
	language = {en},
	urldate = {2025-09-07},
	booktitle = {2025 {IEEE} {Symposium} on {Security} and {Privacy} ({SP})},
	publisher = {IEEE},
	author = {Shioji, Elisa and Meliksetyan, Ani and Simko, Lucy and Watkins, Ryan and Aviv, Adam J. and Cohney, Shaanan},
	month = may,
	year = {2025},
	pages = {2790--2808},
}

@article{cavoukian_privacy_2010,
	title = {Privacy by design: the definitive workshop. {A} foreword by {Ann} {Cavoukian}, {Ph}.{D}},
	volume = {3},
	issn = {1876-0678},
	shorttitle = {Privacy by design},
	url = {http://link.springer.com/10.1007/s12394-010-0062-y},
	doi = {10.1007/s12394-010-0062-y},
	language = {en},
	number = {2},
	urldate = {2025-09-07},
	journal = {Identity in the Information Society},
	author = {Cavoukian, Ann},
	month = aug,
	year = {2010},
	pages = {247--251},
}

@article{burgstahler_what_2021,
	title = {What {Higher} {Education} {Learned} {About} the {Accessibility} of {Online} {Opportunities} {During} a {Pandemic}},
	volume = {21},
	issn = {2158-3595, 2158-3595},
	url = {https://articlegateway.com/index.php/JHETP/article/view/4493},
	doi = {10.33423/jhetp.v21i7.4493},
	language = {en},
	number = {7},
	urldate = {2025-09-07},
	journal = {Journal of Higher Education Theory and Practice},
	author = {Burgstahler, Sheryl},
	month = aug,
	year = {2021},
	pages = {160--170},
}

@inproceedings{bergmans_efficacy_2021,
	address = {Online Streaming, --- Select a Country ---},
	title = {On the {Efficacy} of {Online} {Proctoring} using {Proctorio}:},
	isbn = {978-989-758-502-9},
	shorttitle = {On the {Efficacy} of {Online} {Proctoring} using {Proctorio}},
	url = {https://www.scitepress.org/DigitalLibrary/Link.aspx?doi=10.5220/0010399602790290},
	doi = {10.5220/0010399602790290},
	language = {en},
	urldate = {2025-09-07},
	booktitle = {Proceedings of the 13th {International} {Conference} on {Computer} {Supported} {Education}},
	publisher = {SCITEPRESS - Science and Technology Publications},
	author = {Bergmans, Laura and Bouali, Nacir and Luttikhuis, Marloes and Rensink, Arend},
	year = {2021},
	pages = {279--290},
}

@article{lothes_sitting_2023,
	title = {Sitting {Meditation} and {Mindfulness} {Effects} on {Overall} {Anxiety} and {Test} {Anxiety} {Among} {College} {Students}},
	volume = {17},
	issn = {1751-2271, 1751-228X},
	url = {https://onlinelibrary.wiley.com/doi/10.1111/mbe.12344},
	doi = {10.1111/mbe.12344},
	language = {en},
	number = {1},
	urldate = {2025-09-07},
	journal = {Mind, Brain, and Education},
	author = {Lothes, John E. and Matney, Sara and Naseer, Zayne and Pfyffer, Riley},
	month = feb,
	year = {2023},
	pages = {61--69},
}

@article{antonenko_framework_2017,
	title = {A framework for aligning needs, abilities and affordances to inform design and practice of educational technologies},
	volume = {48},
	copyright = {http://onlinelibrary.wiley.com/termsAndConditions\#vor},
	issn = {0007-1013, 1467-8535},
	url = {https://bera-journals.onlinelibrary.wiley.com/doi/10.1111/bjet.12466},
	doi = {10.1111/bjet.12466},
	language = {en},
	number = {4},
	urldate = {2025-09-07},
	journal = {British Journal of Educational Technology},
	author = {Antonenko, Pavlo D. and Dawson, Kara and Sahay, Shilpa},
	month = jun,
	year = {2017},
	pages = {916--927},
}

@article{schmidt_toward_2024,
	title = {Toward a strengths-based model for designing virtual reality learning experiences for autistic users},
	volume = {28},
	issn = {1362-3613, 1461-7005},
	url = {https://journals.sagepub.com/doi/10.1177/13623613231208579},
	doi = {10.1177/13623613231208579},
	language = {en},
	number = {7},
	urldate = {2025-09-07},
	journal = {Autism},
	author = {Schmidt, Matthew and Newbutt, Nigel and Lee, Minyoung and Lu, Jie and Francois, Marc-Sonley and Antonenko, Pavlo D and Glaser, Noah},
	month = jul,
	year = {2024},
	pages = {1809--1827},
}

@article{wobbrock_ability-based_2018,
	title = {Ability-based design},
	volume = {61},
	issn = {0001-0782, 1557-7317},
	url = {https://dl.acm.org/doi/10.1145/3148051},
	doi = {10.1145/3148051},
	language = {en},
	number = {6},
	urldate = {2025-09-07},
	journal = {Communications of the ACM},
	author = {Wobbrock, Jacob O. and Gajos, Krzysztof Z. and Kane, Shaun K. and Vanderheiden, Gregg C.},
	month = may,
	year = {2018},
	pages = {62--71},
}

@inproceedings{pierres_artificial_2023,
	address = {Winterthur, Switzerland},
	title = {Artificial {Intelligence} in {Higher} {Education}: {Ethical} {Concerns} for {Students} with {Disabilities}},
	volume = {3442},
	isbn = {1613-0073},
	language = {en},
	booktitle = {Proceedings of the 2nd {European} {Workshop} on {Algorithmic} {Fairness}},
	publisher = {CEUR Workshop Proceedings},
	author = {Pierrès, Oriane and Darvishy, Alireza and Christen, Markus},
	month = jun,
	year = {2023},
	pages = {5},
}

@mastersthesis{pilgrim_online_2024,
	address = {Kingston, Ontario, Canada},
	title = {Online {Proctoring} and {Discrimination}: {A} {Critical} {Examination} of {Online} {Proctoring} {Technologies}},
	language = {en},
	school = {Queen’s University},
	author = {Pilgrim, Christina Lynn},
	year = {2024},
}

@inproceedings{mutimukwe_privacy_2023,
	address = {Hawaii},
	title = {Privacy as {Contextual} {Integrity} in {Online} {Proctoring} {Systems} in {Highedr} {Education}: {A} scoping review},
	url = {https://hdl.handle.net/10125/102638},
	language = {en},
	booktitle = {Proceedings of the 56th {Hawaii} {International} {Conference} on {System} {Sciences}},
	author = {Mutimukwe, Chantal and Han, Shengnan and Viberg, Olga and Cerratto-Pargman, Teresa},
	year = {2023},
	pages = {74--82},
}

@inproceedings{burgess_watching_2022,
	address = {Boston, USA},
	title = {Watching the watchers: bias and vulnerability in remote proctoring software},
	isbn = {978-1-939133-31-1},
	url = {https://www.usenix.org/conference/usenixsecurity22/presentation/burgess},
	language = {en},
	booktitle = {Proceedings of the 31st {USENIX} {Security} {Symposium}},
	publisher = {USENIX Association},
	author = {Burgess, Ben and Ginsberg, Avi and Felten, Edward W and Cohney, Shaanan},
	year = {2022},
	pages = {571--588},
}

@inproceedings{slade_learning_2019,
	address = {Tempe AZ USA},
	title = {Learning analytics at the intersections of student trust, disclosure and benefit},
	copyright = {https://www.acm.org/publications/policies/copyright\_policy\#Background},
	isbn = {978-1-4503-6256-6},
	url = {https://dl.acm.org/doi/10.1145/3303772.3303796},
	doi = {10.1145/3303772.3303796},
	language = {en},
	urldate = {2025-09-07},
	booktitle = {Proceedings of the 9th {International} {Conference} on {Learning} {Analytics} \& {Knowledge}},
	publisher = {ACM},
	author = {Slade, Sharon and Prinsloo, Paul and Khalil, Mohammad},
	month = mar,
	year = {2019},
	pages = {235--244},
}

@article{fawns_matter_2022,
	title = {A {Matter} of {Trust}: {Online} {Proctored} {Exams} and the {Integration} of {Technologies} of {Assessment} in {Medical} {Education}},
	volume = {34},
	url = {https://www.taylorfrancis.com/books/9781003263463},
	doi = {10.1080/10401334.2022.2048832},
	language = {en},
	number = {4},
	urldate = {2025-05-13},
	journal = {Teaching and Learning in Medicine},
	author = {Fawns, Tim and Schaepkens, Sven P. C.},
	month = apr,
	year = {2022},
	pages = {444--453},
}

@article{pflugfelder_reddits_2017,
	title = {Reddit’s “{Explain} {Like} {I}’m {Five}”: {Technical} {Descriptions} in the {Wild}},
	volume = {26},
	issn = {1057-2252, 1542-7625},
	shorttitle = {Reddit’s “{Explain} {Like} {I}’m {Five}”},
	url = {https://www.tandfonline.com/doi/full/10.1080/10572252.2016.1257741},
	doi = {10.1080/10572252.2016.1257741},
	language = {en},
	number = {1},
	urldate = {2025-09-05},
	journal = {Technical Communication Quarterly},
	author = {Pflugfelder, Ehren Helmut},
	month = jan,
	year = {2017},
	pages = {25--41},
}

@misc{cambridge_advanced_learners_dictionary__thesaurus_stimming_2013,
	edition = {4},
	title = {Stimming},
	url = {https://dictionary.cambridge.org/us/dictionary/english/stimming},
	journal = {Cambridge Advanced Learner's Dictionary \& Thesaurus},
	publisher = {Cambridge University Press},
	author = {{Cambridge Advanced Learner's Dictionary \& Thesaurus}},
	year = {2013},
}

@misc{dimeo_online_2017,
	title = {Online exam proctoring catches cheaters, raises concerns},
	url = {https://www.insidehighered.com/digital-learning/article/2017/05/10/online-exam-proctoring-catches-cheaters-raises-concerns},
	language = {en},
	urldate = {2025-09-03},
	journal = {Inside Higher Ed},
	author = {Dimeo, Jean},
	month = may,
	year = {2017},
}

@misc{us_department_of_justice_civil_rights_division_ada_2020,
	type = {Government},
	title = {{ADA} {Requirements}: {Testing} {Accommodations}},
	url = {https://www.ada.gov/resources/testing-accommodations/},
	language = {en},
	urldate = {2025-09-02},
	journal = {ADA.gov},
	author = {{U.S. Department of Justice Civil Rights Division}},
	month = feb,
	year = {2020},
}

@incollection{peixoto_when_2022,
	title = {When {Assessment} {Moves} {Home}: {The} {Digital} {Panopticon} in {Higher} {Education}},
	volume = {256},
	booktitle = {Perspectives and {Trends} in {Education} and {Technology}. {Smart} {Innovation}, {Systems} and {Technologies}},
	publisher = {Springer, Singapore},
	author = {Peixoto, Paulo and Almeida, Joana Gomes and Albuquerque, Cristina Pinto},
	year = {2022},
	doi = {10.1007/978-981-16-5063-5_43},
	pages = {517--526},
}

@article{gin_covid-19_2021,
	title = {{COVID}-19 and {Undergraduates} with {Disabilities}: {Challenges} {Resulting} from the {Rapid} {Transition} to {Online} {Course} {Delivery} for {Students} with {Disabilities} in {Undergraduate} {STEM} at {Large}-{Enrollment} {Institutions}},
	volume = {20},
	issn = {1931-7913},
	shorttitle = {{COVID}-19 and {Undergraduates} with {Disabilities}},
	url = {https://www.lifescied.org/doi/10.1187/cbe.21-02-0028},
	doi = {10.1187/cbe.21-02-0028},
	language = {en},
	number = {ar36},
	urldate = {2025-05-13},
	journal = {CBE—Life Sciences Education},
	author = {Gin, Logan E. and Guerrero, Frank A. and Brownell, Sara E. and Cooper, Katelyn M.},
	editor = {Momsen, Jennifer},
	month = sep,
	year = {2021},
	pages = {1--17},
}

@article{hill_accused_2022,
	title = {Accused of {Cheating} by an {Algorithm}, and a {Professor} {She} {Had} {Never} {Met}},
	url = {https://www.nytimes.com/2022/05/27/technology/college-students-cheating-software-honorlock.html},
	journal = {The New York Times},
	author = {Hill, Kashmir},
	month = may,
	year = {2022},
	pages = {9},
}

@article{woldeab_21st_2019,
	title = {21st {Century} {Assessment}: {Online} {Proctoring}, {Test} {Anxiety}, and {Student} {Performance}},
	volume = {34},
	issn = {2292-8588},
	language = {en},
	number = {1},
	journal = {International Journal of E-Learning and Distance Education},
	author = {Woldeab, Daniel and Brothen, Thomas},
	year = {2019},
	pages = {10},
}

@article{patil_how_2020,
	title = {How {It} {Feels} {When} {Software} {Watches} {You} {Take} {Tests}},
	url = {https://www.nytimes.com/2020/09/29/style/testing-schools-proctorio.html},
	language = {en},
	urldate = {2025-08-26},
	journal = {The New York Times},
	author = {Patil, Anushka and Bromwich, Jonah Engel},
	month = sep,
	year = {2020},
	pages = {6},
}

@inproceedings{kuleva_exploring_2024,
	address = {Rezekne, Latvia},
	title = {Exploring the {Efficacy} of {Online} {Proctoring} in  {Online} {Examinations}: {A} {Comprehensive} {Review}},
	volume = {2},
	copyright = {https://creativecommons.org/licenses/by/4.0},
	url = {https://journals.ru.lv/index.php/ETR/article/view/8058},
	doi = {10.17770/etr2024vol2.8058},
	language = {en},
	urldate = {2025-08-24},
	booktitle = {Proceedings of the 15th {International} {Scientific} and {Practical} {Conference}},
	publisher = {Rezekne Academy of Technologies},
	author = {Kuleva, Milena and Miladinov, Ognyan},
	month = jun,
	year = {2024},
	pages = {192--196},
}

@article{johanson_lockdown_2025,
	title = {Lockdown {Lessons}: {How} {Remote} {Proctoring} {Continues} to {Discriminate} {Against} {Disabled} {Students} in a {Post}-{COVID} {Era}},
	volume = {54},
	url = {https://scholarcommons.sc.edu/jled/vol54/iss1/6},
	language = {en},
	number = {1},
	journal = {The Journal of Law and Education},
	author = {Johanson, Abigail},
	year = {2025},
	pages = {6},
}

@article{feathers_proctorio_2021,
	title = {Proctorio {Is} {Using} {Racist} {Algorithms} to {Detect} {Faces}},
	url = {https://www.vice.com/en/article/proctorio-is-using-racist-algorithms-to-detect-faces/},
	language = {en},
	urldate = {2025-08-26},
	journal = {Vice Magazine},
	author = {Feathers, Todd},
	month = apr,
	year = {2021},
	pages = {9},
}

@article{caplan-bricker_is_2021,
	title = {Is {Online} {Test}-{Monitoring} {Here} to {Stay}?},
	url = {https://www.newyorker.com/tech/annals-of-technology/is-online-test-monitoring-here-to-stay},
	language = {en},
	urldate = {2025-08-26},
	journal = {The New Yorker},
	author = {Caplan-Bricker, Nora},
	month = may,
	year = {2021},
	pages = {12},
}

@article{barrio_legal_2022,
	title = {Legal and {Pedagogical} {Issues} with {Online} {Exam} {Proctoring}},
	volume = {13},
	number = {1},
	journal = {European Journal of Law and Technology},
	author = {Barrio, Fernando},
	year = {2022},
	pages = {18},
}

@misc{cambridge_university_press_proctor_nodate,
	title = {Proctor},
	url = {https://dictionary.cambridge.org/us/dictionary/english/proctor},
	urldate = {2025-08-27},
	journal = {Cambridge Advanced Learner's Dictionary \& Thesaurus},
	publisher = {Cambridge University Press},
	author = {{Cambridge University Press}},
}

@article{coghlan_good_2021,
	title = {Good {Proctor} or “{Big} {Brother}”? {Ethics} of {Online} {Exam} {Supervision} {Technologies}},
	volume = {34},
	issn = {2210-5433, 2210-5441},
	shorttitle = {Good {Proctor} or “{Big} {Brother}”?},
	url = {https://link.springer.com/10.1007/s13347-021-00476-1},
	doi = {10.1007/s13347-021-00476-1},
	language = {en},
	number = {4},
	urldate = {2025-05-13},
	journal = {Philosophy \& Technology},
	author = {Coghlan, Simon and Miller, Tim and Paterson, Jeannie},
	month = dec,
	year = {2021},
	pages = {1581--1606},
}

@article{slusky_cybersecurity_2020,
	title = {Cybersecurity of {Online} {Proctoring} {Systems}},
	volume = {29},
	issn = {1941-6679},
	url = {https://scholarworks.lib.csusb.edu/jitim/vol29/iss1/3},
	doi = {10.58729/1941-6679.1445},
	language = {en},
	number = {1},
	urldate = {2025-08-27},
	journal = {Journal of International Technology and Information Management},
	author = {Slusky, Ludwig},
	month = jan,
	year = {2020},
	pages = {56--83},
}

@incollection{ras_online_2018,
	address = {Cham},
	title = {Online {Proctoring} for {Remote} {Examination}: {A} {State} of {Play} in {Higher} {Education} in the {EU}},
	volume = {829},
	isbn = {978-3-319-97806-2 978-3-319-97807-9},
	shorttitle = {Online {Proctoring} for {Remote} {Examination}},
	url = {http://link.springer.com/10.1007/978-3-319-97807-9_8},
	language = {en},
	urldate = {2025-08-27},
	booktitle = {Technology {Enhanced} {Assessment}},
	publisher = {Springer International Publishing},
	author = {Draaijer, Silvester and Jefferies, Amanda and Somers, Gwendoline},
	editor = {Ras, Eric and Guerrero Roldán, Ana Elena},
	year = {2018},
	doi = {10.1007/978-3-319-97807-9_8},
	note = {Series Title: Communications in Computer and Information Science},
	pages = {96--108},
}

@inproceedings{li_massive_2015,
	address = {Vancouver BC Canada},
	title = {Massive {Open} {Online} {Proctor}: {Protecting} the {Credibility} of {MOOCs} certificates},
	isbn = {978-1-4503-2922-4},
	shorttitle = {Massive {Open} {Online} {Proctor}},
	url = {https://dl.acm.org/doi/10.1145/2675133.2675245},
	doi = {10.1145/2675133.2675245},
	language = {en},
	urldate = {2025-08-27},
	booktitle = {Proceedings of the 18th {ACM} {Conference} on {Computer} {Supported} {Cooperative} {Work} \& {Social} {Computing}},
	publisher = {ACM},
	author = {Li, Xuanchong and Chang, Kai-min and Yuan, Yueran and Hauptmann, Alexander},
	month = feb,
	year = {2015},
	pages = {1129--1137},
}

@article{kolski_examining_2018,
	title = {Examining the {Relationship} {Between} {Student} {Test} {Anxiety} and {Webcam} {Based} {Exam} {Proctoring}},
	volume = {21},
	language = {en},
	number = {3},
	journal = {Online Journal of Distance Learning Administration},
	author = {Kolski, Tammi and Weible, Jennifer},
	year = {2018},
	pages = {1--15},
}

@incollection{woldeab_under_2017,
	address = {Minneapolis, MN},
	edition = {1},
	series = {Making a {Difference} {Monograph} {Series}},
	title = {Under the {Watchful} {Eye} of {Online} {Proctoring}},
	isbn = {978-1-946135-37-7},
	language = {en},
	booktitle = {Innovative {Learning} and {Teaching}: {Experiments} {Across} the {Disciplines}},
	publisher = {University of Minnesota Libraries Publishing},
	author = {Woldeab, Daniel and Lindsay, Thomas and Brothen, Thomas},
	month = nov,
	year = {2017},
	pages = {147--160},
}

@inproceedings{dilini_cheating_2021,
	address = {Moratuwa, Sri Lanka},
	title = {Cheating {Detection} in {Browser}-based {Online} {Exams} through {Eye} {Gaze} {Tracking}},
	copyright = {https://ieeexplore.ieee.org/Xplorehelp/downloads/license-information/IEEE.html},
	isbn = {978-1-6654-2000-6},
	url = {https://ieeexplore.ieee.org/document/9657277/},
	doi = {10.1109/ICITR54349.2021.9657277},
	language = {en},
	urldate = {2025-08-27},
	booktitle = {2021 6th {International} {Conference} on {Information} {Technology} {Research} ({ICITR})},
	publisher = {IEEE},
	author = {Dilini, Nimesha and Senaratne, Asara and Yasarathna, Tharindu and Warnajith, Nalin and Seneviratne, Leelanga},
	month = dec,
	year = {2021},
	pages = {1--8},
}

@inproceedings{terpstra_online_2023,
	address = {Hamburg Germany},
	title = {Online {Proctoring}: {Privacy} {Invasion} or {Study} {Alleviation}?: {Discovering} {Acceptability} {Using} {Contextual} {Integrity}},
	isbn = {978-1-4503-9421-5},
	shorttitle = {Online {Proctoring}},
	url = {https://dl.acm.org/doi/10.1145/3544548.3581181},
	doi = {10.1145/3544548.3581181},
	language = {en},
	urldate = {2025-08-24},
	booktitle = {Proceedings of the 2023 {CHI} {Conference} on {Human} {Factors} in {Computing} {Systems}},
	publisher = {ACM},
	author = {Terpstra, Arnout and De Rooij, Alwin and Schouten, Alexander},
	month = apr,
	year = {2023},
	pages = {1--20},
}

@article{james_tertiary_2016,
	title = {Tertiary student attitudes to invigilated, online summative examinations},
	volume = {13},
	issn = {2365-9440},
	url = {https://educationaltechnologyjournal.springeropen.com/articles/10.1186/s41239-016-0015-0},
	doi = {10.1186/s41239-016-0015-0},
	language = {en},
	number = {1},
	urldate = {2025-08-22},
	journal = {International Journal of Educational Technology in Higher Education},
	author = {James, Rosalind},
	month = dec,
	year = {2016},
	pages = {19},
}

@article{arno_state---art_2021,
	title = {State-of-the-{Art} of {Commercial} {Proctoring} {Systems} and {Their} {Use} in {Academic} {Online} {Exams}:},
	volume = {19},
	copyright = {http://creativecommons.org/licenses/by/3.0/deed.en\_US},
	issn = {1539-3100, 1539-3119},
	shorttitle = {State-of-the-{Art} of {Commercial} {Proctoring} {Systems} and {Their} {Use} in {Academic} {Online} {Exams}},
	url = {https://services.igi-global.com/resolvedoi/resolve.aspx?doi=10.4018/IJDET.20210401.oa3},
	doi = {10.4018/IJDET.20210401.oa3},
	language = {en},
	number = {2},
	urldate = {2025-08-22},
	journal = {International Journal of Distance Education Technologies},
	author = {Arnò, Simone and Galassi, Alessandra and Tommasi, Marco and Saggino, Aristide and Vittorini, Pierpaolo},
	month = apr,
	year = {2021},
	pages = {55--76},
}

@article{butler-henderson_systematic_2020,
	title = {A systematic review of online examinations: {A} pedagogical innovation for scalable authentication and integrity},
	volume = {159},
	issn = {03601315},
	shorttitle = {A systematic review of online examinations},
	url = {https://linkinghub.elsevier.com/retrieve/pii/S0360131520302220},
	doi = {10.1016/j.compedu.2020.104024},
	language = {en},
	urldate = {2025-08-22},
	journal = {Computers \& Education},
	author = {Butler-Henderson, Kerryn and Crawford, Joseph},
	month = dec,
	year = {2020},
	pages = {104024},
}

@article{conijn_fear_2022,
	title = {The fear of big brother: {The} potential negative side‐effects of proctored exams},
	volume = {38},
	issn = {0266-4909, 1365-2729},
	shorttitle = {The fear of big brother},
	url = {https://onlinelibrary.wiley.com/doi/10.1111/jcal.12651},
	doi = {10.1111/jcal.12651},
	language = {en},
	number = {6},
	urldate = {2025-08-22},
	journal = {Journal of Computer Assisted Learning},
	author = {Conijn, Rianne and Kleingeld, Ad and Matzat, Uwe and Snijders, Chris},
	month = dec,
	year = {2022},
	pages = {1521--1534},
}

@article{hussein_evaluation_2020,
	title = {An {Evaluation} of {Online} {Proctoring} {Tools}},
	volume = {12},
	copyright = {http://creativecommons.org/licenses/by/4.0/},
	issn = {2304-070X, 1369-9997},
	url = {https://openpraxis.org/article/10.5944/openpraxis.12.4.1113/},
	doi = {10.5944/openpraxis.12.4.1113},
	language = {en},
	number = {4},
	urldate = {2025-08-21},
	journal = {Open Praxis},
	author = {Hussein, Mohammed Juned and Yusuf, Javed and Deb, Arpana Sandhya and Fong, Letila and Naidu, Som},
	month = dec,
	year = {2020},
	pages = {509},
}

@article{hylton_utilizing_2016,
	title = {Utilizing {Webcam}-{Based} {Proctoring} to {Deter} {Misconduct} in {Online} {Exams}},
	volume = {92-93},
	issn = {03601315},
	url = {https://linkinghub.elsevier.com/retrieve/pii/S0360131515300518},
	doi = {10.1016/j.compedu.2015.10.002},
	language = {en},
	urldate = {2024-09-26},
	journal = {Computers \& Education},
	author = {Hylton, Kenrie and Levy, Yair and Dringus, Laurie P.},
	month = jan,
	year = {2016},
	pages = {53--63},
}

@article{atoum_automated_2017,
	title = {Automated {Online} {Exam} {Proctoring}},
	volume = {19},
	copyright = {https://ieeexplore.ieee.org/Xplorehelp/downloads/license-information/IEEE.html},
	issn = {1520-9210, 1941-0077},
	url = {http://ieeexplore.ieee.org/document/7828141/},
	doi = {10.1109/TMM.2017.2656064},
	language = {en},
	number = {7},
	urldate = {2024-09-26},
	journal = {IEEE Transactions on Multimedia},
	author = {Atoum, Yousef and Chen, Liping and Liu, Alex X. and Hsu, Stephen D. H. and Liu, Xiaoming},
	year = {2017},
	pages = {1609--1624},
}

@article{dendir_cheating_2020,
	title = {Cheating in {Online} {Courses}: {Evidence} {From} {Online} {Proctoring}},
	volume = {2},
	issn = {24519588},
	url = {https://linkinghub.elsevier.com/retrieve/pii/S2451958820300336},
	doi = {10.1016/j.chbr.2020.100033},
	language = {en},
	urldate = {2024-09-26},
	journal = {Computers in Human Behavior Reports},
	author = {Dendir, Seife and Maxwell, R. Stockton},
	month = aug,
	year = {2020},
	pages = {100033},
}

@article{douglas_data_2023,
	title = {Data quality in online human-subjects research: {Comparisons} between {MTurk}, {Prolific}, {CloudResearch}, {Qualtrics}, and {SONA}},
	volume = {18},
	issn = {1932-6203},
	shorttitle = {Data quality in online human-subjects research},
	url = {https://dx.plos.org/10.1371/journal.pone.0279720},
	doi = {10.1371/journal.pone.0279720},
	language = {en},
	number = {3},
	urldate = {2025-06-16},
	journal = {PLOS ONE},
	author = {Douglas, Benjamin D. and Ewell, Patrick J. and Brauer, Markus},
	editor = {Hallam, Jeffrey S.},
	month = mar,
	year = {2023},
	pages = {e0279720},
}

@article{stoycheff_privacy_2019,
	title = {Privacy and the {Panopticon}: {Online} mass surveillance’s deterrence and chilling effects},
	volume = {21},
	issn = {1461-4448, 1461-7315},
	shorttitle = {Privacy and the {Panopticon}},
	url = {https://journals.sagepub.com/doi/10.1177/1461444818801317},
	doi = {10.1177/1461444818801317},
	language = {en},
	number = {3},
	urldate = {2025-06-08},
	journal = {New Media \& Society},
	author = {Stoycheff, Elizabeth and Liu, Juan and Xu, Kai and Wibowo, Kunto},
	month = mar,
	year = {2019},
	pages = {602--619},
}

@book{braun_thematic_2022,
	address = {Los Angeles},
	title = {Thematic {Analysis}: {A} {Practical} {Guide}},
	publisher = {Sage},
	author = {Braun, Virginia and Clarke, Victoria},
	year = {2022},
}

@article{richards_intellectual_2008,
	title = {Intellectual {Privacy}},
	volume = {87},
	number = {387},
	journal = {Texas Law Review},
	author = {Richards, Neil M.},
	month = dec,
	year = {2008},
	pages = {387--446},
}

@phdthesis{carter_attitudes_2024,
	type = {Dissertation},
	title = {Attitudes {When} {Taking} {Assessments} {Using} {Proctoring} {Software} and {Academic} {Honesty} {Between} {Students} with {Disabilities} and {General} {Population} {Students}},
	language = {en},
	school = {WinonaStateUniversity},
	author = {Carter, Erin},
	month = apr,
	year = {2024},
}

@incollection{reutlinger_its_2022,
	title = {“{It}’s {Meant} to {Be} a {Hazing} {Process}”: {Deciphering} {Ableism} {Surrounding} {Academic} {Accommodations}},
	isbn = {978-90-04-51269-6 978-90-04-51270-2},
	url = {https://brill.com/view/title/62070},
	language = {en},
	urldate = {2025-05-10},
	booktitle = {Redefining {Disability}},
	publisher = {BRILL},
	author = {Reutlinger, Corey},
	month = feb,
	year = {2022},
	doi = {10.1163/9789004512702},
	pages = {240--255},
}

@article{lett_impact_2020,
	title = {Impact of ableist microaggressions on university students with self-identified disabilities},
	volume = {35},
	issn = {0968-7599, 1360-0508},
	url = {https://www.tandfonline.com/doi/full/10.1080/09687599.2019.1680344},
	doi = {10.1080/09687599.2019.1680344},
	language = {en},
	number = {9},
	urldate = {2025-05-13},
	journal = {Disability \& Society},
	author = {Lett, Kayla and Tamaian, Andreea and Klest, Bridget},
	month = oct,
	year = {2020},
	pages = {1441--1456},
}

@article{harwell_cheating-detection_2020,
	title = {Cheating-{Detection} {Companies} {Made} {Millions} {During} the {Pandemic}. {Now} {Students} {Are} {Fighting} back},
	url = {https://www.washingtonpost.com/technology/2020/11/12/test-monitoring-student-revolt/},
	language = {en},
	urldate = {2025-05-13},
	journal = {The Washington Post},
	author = {Harwell, Drew},
	month = nov,
	year = {2020},
	doi = {10.1201/9781003278290-60},
}

@article{pokorny_out_2023,
	title = {“{Out} of my control”: science undergraduates report mental health concerns and inconsistent conditions when using remote proctoring software},
	volume = {19},
	issn = {1833-2595},
	shorttitle = {“{Out} of my control”},
	url = {https://edintegrity.biomedcentral.com/articles/10.1007/s40979-023-00141-4},
	doi = {10.1007/s40979-023-00141-4},
	language = {en},
	number = {1},
	urldate = {2025-05-13},
	journal = {International Journal for Educational Integrity},
	author = {Pokorny, Annika and Ballen, Cissy J. and Drake, Abby Grace and Driessen, Emily P. and Fagbodun, Sheritta and Gibbens, Brian and Henning, Jeremiah A. and McCoy, Sophie J. and Thompson, Seth K. and Willis, Charles G. and Lane, A. Kelly},
	month = nov,
	year = {2023},
	pages = {22},
}

@article{grimes_university_2019,
	title = {University student perspectives on institutional non-disclosure of disability and learning challenges: reasons for staying invisible},
	volume = {23},
	issn = {1360-3116, 1464-5173},
	shorttitle = {University student perspectives on institutional non-disclosure of disability and learning challenges},
	url = {https://www.tandfonline.com/doi/full/10.1080/13603116.2018.1442507},
	doi = {10.1080/13603116.2018.1442507},
	language = {en},
	number = {6},
	urldate = {2025-05-10},
	journal = {International Journal of Inclusive Education},
	author = {Grimes, Susan and Southgate, Erica and Scevak, Jill and Buchanan, Rachel},
	month = jun,
	year = {2019},
	pages = {639--655},
}

@article{roslin_vitriolic_2021,
	title = {Vitriolic {Verification}: {Accommodations}, {Overbroad} {Medical} {Record} {Requests}, and {Procedural} {Ableism} in {Higher} {Education}},
	volume = {47},
	copyright = {https://www.cambridge.org/core/terms},
	issn = {0098-8588, 2375-835X},
	shorttitle = {Vitriolic {Verification}},
	url = {https://www.cambridge.org/core/product/identifier/S0098858821000083/type/journal_article},
	doi = {10.1017/amj.2021.8},
	language = {en},
	number = {1},
	urldate = {2025-05-10},
	journal = {American Journal of Law \& Medicine},
	author = {Roslin, Tara},
	month = mar,
	year = {2021},
	pages = {109--130},
}

@article{brown_ableism_2018,
	title = {Ableism in academia: where are the disabled and ill academics?},
	volume = {33},
	issn = {0968-7599, 1360-0508},
	shorttitle = {Ableism in academia},
	url = {https://www.tandfonline.com/doi/full/10.1080/09687599.2018.1455627},
	doi = {10.1080/09687599.2018.1455627},
	language = {en},
	number = {6},
	urldate = {2025-05-09},
	journal = {Disability \& Society},
	author = {Brown, Nicole and Leigh, Jennifer},
	month = jul,
	year = {2018},
	pages = {985--989},
}

@article{braun_saturate_2021,
	title = {To saturate or not to saturate? {Questioning} data saturation as a useful concept for thematic analysis and sample-size rationales},
	volume = {13},
	issn = {2159-676X, 2159-6778},
	shorttitle = {To saturate or not to saturate?},
	url = {https://www.tandfonline.com/doi/full/10.1080/2159676X.2019.1704846},
	doi = {10.1080/2159676X.2019.1704846},
	language = {en},
	number = {2},
	urldate = {2025-04-22},
	journal = {Qualitative Research in Sport, Exercise and Health},
	author = {Braun, Virginia and Clarke, Victoria},
	month = mar,
	year = {2021},
	pages = {201--216},
}

@inproceedings{prinsloo_student_2015,
	address = {Poughkeepsie New York},
	title = {Student privacy self-management: implications for learning analytics},
	isbn = {978-1-4503-3417-4},
	shorttitle = {Student privacy self-management},
	url = {https://dl.acm.org/doi/10.1145/2723576.2723585},
	doi = {10.1145/2723576.2723585},
	language = {en},
	urldate = {2025-02-03},
	booktitle = {Proceedings of the {Fifth} {International} {Conference} on {Learning} {Analytics} {And} {Knowledge}},
	publisher = {ACM},
	author = {Prinsloo, Paul and Slade, Sharon},
	month = mar,
	year = {2015},
	pages = {83--92},
}

@article{yoder-himes_racial_2022,
	title = {Racial, skin tone, and sex disparities in automated proctoring software},
	volume = {7},
	issn = {2504-284X},
	url = {https://www.frontiersin.org/articles/10.3389/feduc.2022.881449/full},
	doi = {10.3389/feduc.2022.881449},
	language = {en},
	urldate = {2024-10-03},
	journal = {Frontiers in Education},
	author = {Yoder-Himes, Deborah R. and Asif, Alina and Kinney, Kaelin and Brandt, Tiffany J. and Cecil, Rhiannon E. and Himes, Paul R. and Cashon, Cara and Hopp, Rachel M. P. and Ross, Edna},
	month = sep,
	year = {2022},
	pages = {881449},
}

@article{slade_learning_2013,
	title = {Learning {Analytics}: {Ethical} {Issues} and {Dilemmas}},
	volume = {57},
	issn = {0002-7642},
	shorttitle = {Learning {Analytics}},
	url = {https://doi.org/10.1177/0002764213479366},
	doi = {10.1177/0002764213479366},
	language = {en},
	number = {10},
	urldate = {2024-09-12},
	journal = {American Behavioral Scientist},
	author = {Slade, Sharon and Prinsloo, Paul},
	month = oct,
	year = {2013},
	note = {Publisher: SAGE Publications Inc},
	pages = {1510--1529},
}

@inproceedings{kwapisz_privacy_2024,
	address = {Honolulu, HI, USA},
	title = {Privacy {Concerns} of {Student} {Data} {Shared} with {Instructors} in an {Online} {Learning} {Management} {System}},
	doi = {10.1145/3613904.3642914},
	language = {en},
	booktitle = {Proceedings of the {CHI} {Conference} on {Human} {Factors} in {Computing} {Systems} ({CHI} ’24)},
	publisher = {ACM},
	author = {Kwapisz, Monika Blue and Kohli, Avanya and Rajivan, Prashanth},
	month = may,
	year = {2024},
	pages = {16},
}

@inproceedings{kelley_nutrition_2009,
	address = {Mountain View, California, USA},
	title = {A "{Nutrition} {Label}" for {Privacy}},
	doi = {10.1145/1572532.1572538},
	booktitle = {{SOUPS} '09: {Proceedings} of the 5th {Symposium} on {Usable} {Privacy} and {Security}},
	publisher = {Association for Computing Machinery},
	author = {Kelley, Patrick Gage and Bresee, Joanna and Cranor, Lorrie Faith and Reeder, Robert W.},
	month = jul,
	year = {2009},
	pages = {1--12},
}

@inproceedings{khan_exploring_2023,
	address = {Ahaheim, CA, USA},
	title = {Exploring {Privacy} and {Security} {Concerns} of {EdTech} {Users}: {A} {Qualitative} {Analysis} of {User} {Written} {Reviews}},
	language = {en},
	booktitle = {{USENIX} {Symposium} on {Usable} {Privacy} and {Security} ({SOUPS}) 2023},
	publisher = {USENIX Association},
	author = {Khan, Waqar Hassan and Pranto, Protik Bose and Yang, Tianyi and Hasan, Rakibul},
	month = aug,
	year = {2023},
	pages = {7},
}

@inproceedings{ul_haque_nuanced_2023,
	address = {Hamburg Germany},
	title = {The {Nuanced} {Nature} of {Trust} and {Privacy} {Control} {Adoption} in the {Context} of {Google}},
	isbn = {978-1-4503-9421-5},
	url = {https://dl.acm.org/doi/10.1145/3544548.3581387},
	doi = {10.1145/3544548.3581387},
	language = {en},
	urldate = {2024-09-07},
	booktitle = {Proceedings of the 2023 {CHI} {Conference} on {Human} {Factors} in {Computing} {Systems}},
	publisher = {ACM},
	author = {Ul Haque, Ehsan and Khan, Mohammad Maifi Hasan and Fahim, Md Abdullah Al},
	month = apr,
	year = {2023},
	pages = {1--23},
}

@article{oates_turtles_2018,
	title = {Turtles, {Locks}, and {Bathrooms}: {Understanding} {Mental} {Models} of {Privacy} {Through} {Illustration}},
	volume = {2018},
	issn = {2299-0984},
	shorttitle = {Turtles, {Locks}, and {Bathrooms}},
	url = {https://petsymposium.org/popets/2018/popets-2018-0029.php},
	doi = {10.1515/popets-2018-0029},
	language = {en},
	number = {4},
	urldate = {2024-03-12},
	journal = {Proceedings on Privacy Enhancing Technologies},
	author = {Oates, Maggie and Ahmadullah, Yama and Marsh, Abigail and Swoopes, Chelse and Zhang, Shikun and Balebako, Rebecca and Cranor, Lorrie Faith},
	month = oct,
	year = {2018},
	pages = {5--32},
}

@article{grayson_education_1978,
	title = {Education, technology, and individual privacy},
	volume = {26},
	issn = {0148-5806},
	url = {https://link.springer.com/10.1007/BF02766604},
	doi = {10.1007/BF02766604},
	language = {en},
	number = {3},
	urldate = {2024-02-09},
	journal = {Educational technology research and development},
	author = {Grayson, Lawrence P.},
	month = sep,
	year = {1978},
	pages = {195--206},
}

@inproceedings{kang_my_2015,
	title = {"{My} {Data} {Just} {Goes} {Everywhere}:" {User} {Mental} {Models} of the {Internet} and {Implications} for {Privacy} and {Security}},
	booktitle = {Symposium on {Usable} {Privacy} and {Security} ({SOUPS})},
	author = {Kang, Ruogu and Dabbish, Laura and Fruchter, Nathaniel and Kiesler, Sara},
	year = {2015},
	pages = {39--52},
}

@book{charmaz_constructing_2006,
	address = {London ; Thousand Oaks, Calif},
	title = {Constructing grounded theory},
	isbn = {978-0-7619-7352-2 978-0-7619-7353-9},
	language = {en},
	publisher = {Sage Publications},
	author = {Charmaz, Kathy},
	year = {2006},
}

@book{foucault_powerknowledge_1980,
	address = {New York},
	edition = {1st},
	title = {Power/knowledge: selected interviews and other writings, 1972-1977},
	isbn = {978-0-394-51357-7 978-0-394-73954-0},
	shorttitle = {Power/knowledge},
	language = {en},
	publisher = {Pantheon Books},
	author = {Foucault, Michel},
	translator = {Gordon, Colin and Marshall, Leo and Mepham, John and Soper, Kate},
	year = {1980},
}

@article{selwyn_necessary_2023,
	title = {A necessary evil? {The} rise of online exam proctoring in {Australian} universities},
	volume = {186},
	number = {1},
	journal = {Media International Australia},
	author = {Selwyn, Neil and O’Neill, Chris and Smith, Gavin and Andrejevic, Mark and Gu, Xin},
	year = {2023},
	note = {Publisher: SAGE Publications Sage UK: London, England},
	pages = {149--164},
}

@incollection{jiang_data_2023,
	address = {Cham},
	title = {Data {Privacy} in {Learning} {Management} {Systems}: {Perceptions} of {Students}, {Faculty}, and {Administrative} {Staff}},
	volume = {14060},
	isbn = {978-3-031-48059-1 978-3-031-48060-7},
	shorttitle = {Data {Privacy} in {Learning} {Management} {Systems}},
	url = {https://link.springer.com/10.1007/978-3-031-48060-7_8},
	language = {en},
	urldate = {2024-02-10},
	booktitle = {{HCI} {International} 2023 – {Late} {Breaking} {Papers}},
	publisher = {Springer Nature Switzerland},
	author = {Jiang, Jialun Aaron and Robledo Yamamoto, Fujiko and Nagy, Vaughan and Zander, Madelyn and Barker, Lecia},
	year = {2023},
	doi = {10.1007/978-3-031-48060-7_8},
	note = {Series Title: Lecture Notes in Computer Science},
	pages = {100--115},
}

@book{foucault_discipline_1977,
	address = {New York},
	title = {Discipline and {Punish}: {The} {Birth} of the {Prison}},
	language = {en},
	publisher = {Vintage},
	author = {Foucault, Michel},
	translator = {Sheridan, Alan},
	year = {1977},
}

@article{khalil_nexus_2022,
	title = {In the nexus of integrity and surveillance: {Proctoring} (re)considered},
	volume = {38},
	issn = {0266-4909, 1365-2729},
	shorttitle = {In the nexus of integrity and surveillance},
	url = {https://onlinelibrary.wiley.com/doi/10.1111/jcal.12713},
	doi = {10.1111/jcal.12713},
	language = {en},
	number = {6},
	urldate = {2024-02-09},
	journal = {Journal of Computer Assisted Learning},
	author = {Khalil, Mohammad and Prinsloo, Paul and Slade, Sharon},
	month = dec,
	year = {2022},
	pages = {1589--1602},
}

\appendix

\section{Appendix}

\section{Interview guide} \label{guide}

The interview started with rapport-building questions about how the participant prepares for exams, how and where they take exams, how they feel before and during exams, and what their accommodations are. Then, we asked orienting questions about when, where, and how the participant takes online proctored exams. Then we moved into the main topics of the interview: mental models of privacy, misrepresentation, ableism, justification, and future-facing questions. The phrasing and order of the questions differed per the intensive interviewing style.

\begin{itemize}
    \item \textbf{Mental models of privacy:} Understand how students view their privacy on the OPS and how they view their privacy generally
    
    Example questions:
    \begin{itemize}
        \item How would you describe online proctoring to a 5-year-old?
        \begin{itemize}
            \item Follow up on mentions of watching and monitoring. Ask how they feel about watching.
        \end{itemize}
        \item How would you define personal privacy?
        \begin{itemize}
            \item Follow up on mentions of thought and belief, private spaces, intellectual exploration, confidential communications
            \item How would you apply your definition of privacy to a testing setting?
        \end{itemize}
        \item What type of data do you think is being viewed or collected by the proctoring technology?
    \end{itemize} 
    \item \textbf{Misrepresentation:} Understand how feelings about online proctoring are related to being accused of cheating or seen as suspicious.

    Example questions:
    \begin{itemize}
        \item Have you ever felt like you may be accused of cheating on test proctoring technology?
        \begin{itemize}
            \item Probe on actual experiences of being flagged or accused.
            \item Follow up on emotions during the experience.
        \end{itemize}
        \item What do you do during the exam to avoid being accused of cheating?
        \begin{itemize}
            \item Follow up on the effect on the test-taking experience.
        \end{itemize}
    \end{itemize}
    
    \item \textbf{Ableism:} Understand how the participant's disability affects their experience with the OPS.

    Example questions:
    \begin{itemize}
        \item Have you ever felt like you couldn’t meet a need you had because of the online proctoring?
        \begin{itemize}
            \item Follow up on going to the bathroom, taking medication, using accommodations, and accessibility in the OPS.
        \end{itemize}
    \end{itemize}

    \item \textbf{Justification:} Understand the reasons the participant gives for why they believe OPS should or shouldn't be used.

    Example questions:
    \begin{itemize}
        \item What is the role of proctoring software in maintaining academic integrity?
        \begin{itemize}
            \item Follow up on what aspects and features are necessary, important, or effective.
        \end{itemize}
    \end{itemize}
    
    \item \textbf{Future facing:} Understand students' desires for OPS and testing in general.

    Example questions:
    \begin{itemize}
        \item In an ideal world, how would you change online proctoring to be most accommodating of you?
        \item In an ideal world, how would online testing best serve you and your education? 
    \end{itemize}
\end{itemize}

\section{Codebook} \label{codebook}

\textbf{Accommodations affect privacy:} Participants report that they have more privacy because they take tests at a testing center.

\textbf{Accessible future:} This code compiles participants' various visions of how OP and online testing can be more accessible and serve them better in an ideal world. This code captures suggested improvements to the current system, as well as offering alternatives that make OP more accessible and efficient for all stakeholders.

\textbf{Assumed surveillance:} Participants say they assume someone is either watching recordings or watching them through the live camera during the exam, even if it is unclear to them if there is a proctor watching.

\textbf{Cheating prevention:} Participants say the OP is there to prevent cheating, plagiarism, or AI use, and to encourage honesty.

\textbf{Control:} Participants defining privacy as control or discussing discomfort with a lack of control

\textbf{Creepy cameras:} Participant referring to discomfort around cameras, not just neutral discussion of cameras, especially calling the cameras "creepy," "odd," "weird," "unsettling"

\textbf{Disorganized accommodations:} Participants say that implementing their accommodations is difficult or disorganized due to limitations of the OP, factors related to the instructor, or institutional policies. For example: Time is cut off by OP, the professor is unaccustomed to implementing accommodations, the professor doesn't coordinate accommodations, the institution requires extra steps for using accommodations with OP

\textbf{Different than others:} The participant sets themselves apart from other students because they have more or less strict privacy boundaries.

\textbf{Doubting surveillance:} Participants wonder if anyone actually looks at the cameras or recordings, especially with the implication that there is doubt if monitoring is actually happening.

\textbf{Fairness:} Participants say testing should be based on merit and effort. People with disabilities should be allowed accommodations, but not special privileges. Participants sometimes say they are being mindful of only using accommodations as necessary.

\textbf{Gossip:} Participants wanting to avoid being talked about for behavior during the test, observed by the OP, or their accommodations being discussed unnecessarily.

\textbf{Humanizing self:} Participants describing engaging in behavior to seem more relatable to the OP. Participants want to be seen as a person, especially in relation to an automated system. Participants may need to explain their accommodations or actions.

\textbf{Implementation discrepancy:} Participants recounting discrepancies between cameras and in-person proctors in various rooms, various types of OP in various classes, and the type of proctoring being implemented irrationally. Participants say the OP is arbitrary, random. or optional because it is not consistently required for every class or by every instructor. Different versions of OP software are used arbitrarily (e.g., LockDown Browser, HonorLock, ExamSoft)

\textbf{In-person proctor:}	Participants comparing OP to in-person proctoring or OP being used in addition to in-person proctoring.

\textbf{Limitations of online proctoring:}	Participants say OP can only see limited information, such as what the camera sees of the student is limited.

\textbf{Monitored and watched:}	Participants using the words "monitored" or "watched" additionally "staring" when describing OP.

\textbf{Movement:} Participants recount getting in trouble or the OP system flagging for fidgeting, standing up and down, looking away, or leaning into the screen.

\textbf{Multiple contexts:}	Participants compare how surveillance is used in various contexts, such as policing vs proctoring, societal vs testing settings. Participants sometimes say monitoring is okay in some contexts but not others.

\textbf{Necessary:}	Participants saying everyone uses online proctoring. It is needed for online learning. They have no choice but to use OP.

\textbf{Negative emotional experience:}	Participants feeling frustrated or flustered, feeling added pressure, or anxiety due to some aspect of OP or being watched.

\textbf{Negative effect on focus:} Participants report that being watched caused negative effects on the test-taking experience, especially focus and performance.

\textbf{Negative tech experience:} Participants report a negative testing experience due to the actual OP software functionality. For example: Wi-fi issues, failing OP software, and inaccessible help with OP software during testing.

\textbf{No code:} Agreement with interviewer, pleasantries, possible incorrect structural coding

\textbf{One-sided proctoring:} Participants are unable to reach the proctor to resolve a situation or defend themselves from potential accusations, which may lead to feeling the proctoring is intense because the proctor is inaccessible.

\textbf{Online proctoring requirements:} Participants recounting the steps to using OP during tests, including what tech was used and how to use it.

\textbf{Overthinking:} Participants describe thinking too much about behaviors and how they will be perceived by the OP. The participant is not explicit about being perceived as cheating or noncompliant, but they are worried about being perceived as suspicious or judged by a remote human proctor.

\textbf{Permissive:} The participant not minding, being content, or being unbothered by monitoring, in-person or online. This code is a collection of reasons someone might feel permissive, such as: "Google already has everything mindset," wanting to have a simple testing experience, or upholding a safe society. Also includes forgetting about the OP while focusing on the exam.

\textbf{Positive relationships:} Participants describe the staff in the disability office, instructor, or the OP staff as nice and accommodating. They may have more trust in their school or the OP due to positive relationships and not having any problems with the school, staff, or OP.

\textbf{Preferred surveillance:} Participants express a preference based on various traits of the preferred proctoring. For example: webcam vs "aerial" view or second camera view, known vs unknown proctor, in-person proctor vs OP, human vs AI

\textbf{Preventative action:} Participants taking a deliberate, additional action to prevent a negative outcome. For example: avoid being talked about or perceived negatively, trying to fit in or not stand out, trying to influence the proctor, or manipulating the OP itself

\textbf{Private about disability:} Participants not wanting to disclose their disability or accommodations to stakeholders of the OP and testing process, such as proctors, testing center staff, or other students.

\textbf{Proctor's capabilities:} Participants speculating (because there is a lack of clarity) about what the proctor has control over and what the proctor has access to

\textbf{Protecting others:}	Participants are mindful of kids and family members being captured in the OP monitoring because they do not want them to have privacy violations. Additionally, participants want parental control over their children's privacy.

\textbf{Recording and stress:} Participants say they do not want to be monitored in a stressed-out state, and OP adds to the stress of test-taking because of being on camera.

\textbf{Rushing:} Participants report that their dislike of the OP makes them rush through the test.

\textbf{Scared of misunderstanding:} Participants worry that their actions will be perceived as cheating or noncompliant, which may include being self-conscious about their actions, but this code captures explicit mentions of the concern about being seen as cheating or noncompliant. Includes “not taking a risk.”

\textbf{Self-conscious:} Participants report that they regulate and limit their true behavior and actions during the test because they believe someone might be watching them. This may include language such as “self-cognizant”

\textbf{Societal targeting:} Participants feel the need to censor because they are afraid of targeting from the OP based on an aspect of identity, such as political opinions or identity that has been politicized and policed (ex, race). This includes feeling that there could be unfair and nonconsensual advantages or disadvantages given to students based on race or other characteristics.

\textbf{Stakes of online proctoring:} Participants reporting that their performance on test is related to high-stakes future issues, like economic well-being and good grades

\textbf{Testing equity:} Participants state that different needs need to be met to "equal the playing field." For example, an environment free of OP for people who are disproportionately stressed by OP.

\textbf{Trade-off:}	Participant understands that they must be online proctored in exchange for accessible remote test-taking. They may report valuing OP to support better education. For example, proctoring allows for teachers or advisors to reach out to students in regards to areas of struggle during test-taking and offer solutions to be more accommodating and helpful

\textbf{Transparency:} Participants want transparency about the OP because they are unaware of what OP is happening, who is witnessing an emotional reaction, or what features are available in the OP. May include mentions of terms and conditions or the effects of a lack of transparency, like having a lot to remember about the OP.

\textbf{Treated like a child:} Participants report being infantilized by OP or the process of accessing their accommodations. They may feel that they are monitored more strictly than their peers without accommodations. The intensity of the monitoring causes participants to feel like they are losing agency.

\textbf{Trust:}	Participants feel a lack of trust due to being intensely monitored, especially in the "hyper-structured environment" or a test.

\textbf{Unclear mechanisms:} Participants say they don't understand how the OP works on a technical level or what features are available.

\textbf{Understand cheating:} Participants understand why people cheat, feeling like people are always going to cheat, or understanding that instructors are also aware of cheating and AI use.

\textbf{Unnecessary:} Proctoring is being used when not necessary, for example, workers have resources, so why can't students? Surveillance is picking up info that is not part of the test-taking experience. Can also be framed as "doing too much" too much time, resources, vigilance

\textbf{Used to it:} Participants saying they are used to proctoring or knowing when the OP would flag them.

\textbf{Values privacy:} Participants describe themselves as having privacy values, like defining themselves as private people, questioning the ethics of OP, and giving opinions on privacy being lacking throughout online platforms, or generally being a privacy-minded person

\textbf{Vulnerability:}	Definition of privacy as vulnerability, what is vulnerable for people? Types of information people are opposed to. Ex. Data, where they live, and non-consensual use of data

\textbf{Weird:}	Participants expressing OP is "weird," "strange," unsettling, or unnatural

\section{Image descriptions}

\subsection{Image description for Figure~\ref{fig: anxiety}} \label{image des: anxiety} 
The box on the left contains the codes under surveillance-related causes of anxiety: assumed surveillance, gossip, monitored and watched, one-sided proctoring, private about disability, proctor's capabilities, recording and stress, treated like a child, trust, and weird.
And implementation-related causes of anxiety: implementation discrepancy, negative tech experience, and stakes of online proctoring. And arrow points to the anxiety and frustration box that contains the code, negative emotional experience. An arrow points out of Anxiety and frustration to Fear of misrepresentation, which contains the codes: scared of misunderstanding, self-conscious, and overthinging. From there, an arrow points to a box titled Cognitive load increase, which contains: humanizing self, movement, preventative action, negative effect on focus, and rushing. An arrow points back to Anxiety and frustration, demonstrating the cycle described by the theme in Sections~\ref{theme: anxiety}.

\subsection{Image description for Figure~\ref{fig: compromise}} \label{image des: compromise}

Figure~\ref{fig: compromise} depicts a balance scale with benefits on one side and disadvantages on the other side, with a fulcrum labeled ``compromise.'' On the benefits side, the codes stacked on top of the scale are accommodations affect privacy, cheating prevention, different than others, multiple contexts, necessary, permissive, and used to it. On the disadvantages side, the codes are: disorganized accommodations, limitations of online proctoring, and unnecessary. The trade-off code sits above the fulcrum.

\subsection{Image description for Figure~\ref{fig: wordcloud}} \label{image des: word cloud}

Figure~\ref{fig: wordcloud} is a word cloud created based on the following table. The vulnerability factors in the first column appear larger in the word cloud if their frequency (second column) is larger.

\begin{table}[h]
\caption{Data basis of word cloud}
\begin{tabular}{ll}
\toprule
Vulnerability factor & Frequency \\
\midrule
home                                  & 6 \\
data storage                          & 5 \\
personally identifiable   information & 4 \\
stranger                              & 4 \\
background programs                   & 3 \\
leak or hack                          & 3 \\
location                              & 3 \\
health condition                      & 2 \\
identification number                 & 2 \\
images                                & 2 \\
messages                              & 2 \\
access to you                         & 1 \\
dark web                              & 1 \\
facial recognition                    & 1 \\
intellectual property                 & 1 \\
personal life                         & 1 \\
robot                                 & 1 \\
selling data                          & 1 \\
spam                                  & 1 \\
unique information                    & 1 \\
\bottomrule
\end{tabular}
\end{table}

\section{Demographic tables} \label{demographics}

\begin{table*}[h]
\caption{Demographics}
\centering
\begin{tabular}{lcr}
\toprule
Demographics & All students \textit{n} = 16 & \% (\textit{n}) \\
\midrule
Disability types & & \\
 \quad Mental health/psychological disability (e.g., anxiety, depression, PTSD) & & 94\% (15)\\
 \quad Learning disability (e.g. dyslexia, ADHD) & &44\% (7)\\
 \quad Physical disability (e.g., cerebral palsy, spina bifida, dwarfism) & & 44\% (7)\\
 \quad Chronic health condition (e.g., cancer, diabetes, multiple sclerosis) & & 38\% (6)\\
 \quad Visual loss (e.g. blind) & & 19\% (3)\\
 \quad Hearing loss (e.g. D/deaf) & & 13\% (2)\\
 Gender& &\\
 \quad Male & & 75\% (12)\\
 \quad Female & & 44\% (7)\\
 \quad Declined to answer & & 6\% (1)\\
 \quad Nonbinary & & 6\% (1)\\
 \quad Queer & & 6\% (1)\\
 \quad Two spirited & & 6\% (1)\\
 Identify as LGBTQ? & &\\
 \quad Yes & & 50\% (8)\\
 \quad No & & 50\% (8)\\
 Race/Ethnicity& &\\
 \quad White & & 44\% (7)\\
 \quad Hispanic & & 13\% (2)\\
 \quad Alaskan Native & & 6\% (1)\\
 \quad Black & & 6\% (1)\\
 \quad Black American and Mexican & & 6\% (1)\\
 \quad Mixed & & 6\% (1)\\
 \quad South Asian& & 6\% (1)\\
 \quad Southeast Asian & & 6\% (1)\\
 \quad White/Black & & 6\% (1)\\
 Are you a first-generation college student? & &\\
 \quad No & & 63\% (10)\\
 \quad Yes & & 38\% (6)\\
 Degree Program & &\\
 \quad Undergraduate & & 69\% (11)\\
 \quad Graduate& & 31\% (5)\\
 Undergraduate Years in School& &\\
 \quad 1 year or less (first-year student)& & 45\% (5)\\
 \quad 2 years (sophomore)& & 18\% (2)\\
 \quad 3 years (junior)& & 18\% (2)\\
 \quad 4 years (senior)& & 9\% (1)\\
\quad 5 years or more& & 9\% (1)\\
\bottomrule
\end{tabular}
\end{table*}

\begin{table*}[h]
  \caption{Accommodations}
  \label{tab: accommodations}
  \begin{tabular}{lp{3.5in}c}
    \toprule
    Accommodation & Definition & Frequency\\
    \midrule 
    Reader & Student can have an aide read written text out loud or use screen reader technology & 5 \\
    Extra time & Student has extra time, such as time and a half when taking the exam & 5 \\
    Testing center & Student takes exam in a testing center & 4 \\
    Setting change & Student can request to take their test in a different setting, such as taking it in a different room, at home, or even a smaller room & 4 \\
    Extension & Student can request to take their test at a later date or turn in assignments at a later time & 4 \\
    Notes & Student is allowed to have notes or drawings as assistance during testing & 3 \\
    Break & Student is allowed to have breaks & 3 \\
    Reduced distractions & Student takes test in a reduced distraction environment, such as a quieter room & 2 \\
    Move around & Student is allowed to get up, move around, or leave the room & 2 \\
    Magnifier & Student uses software that makes texts and images on the screen larger & 2 \\
    Light & Student has accommodations that apply to the lighting of their room or screen & 2 \\
    Dictation & Student uses software that dictates speech to text & 2 \\
    Writer & Student can have someone write out their answers for them  & 1 \\
    Printout & Student can request printed versions of online materials & 1 \\
    OCR & Student uses software that analyzes images and converts them into text & 1 \\
    Food and water & Student can bring in food and water during the test & 1 \\
    Calculator & Student is allowed a calculator while taking exams & 1 \\
    \bottomrule
    \end{tabular}
\end{table*}

\end{document}